\documentclass[12pt]{article}
\usepackage{latexsym}
\usepackage{graphics}
\usepackage{epsfig}
\usepackage{booktabs}
\usepackage{epstopdf}
\usepackage{amsmath}
\usepackage{cite}
\usepackage{hyperref}
\usepackage[toc,page]{appendix}
\usepackage{amssymb}

\begin{document}

\begin{center}
\textbf{\Large 
Cluster and statistical analysis of spatial earthquake patterns  in the South Caucasus region 
} \vspace{0.5 cm}

Sergii Skurativskyi$^1$\footnote{e-mail: \url{skurserg@gmail.com}},  Sergiy Mykulyak$^1$,  Yuliya Semenova $^{1,2}$,   Kateryna Skurativska $^3$  \vspace{0.5 cm}

$^1$ S.I.Subbotin Institute of Geophysics of the National Academy of Science of Ukraine, Kyiv, Ukraine

 $^2$ DGFI-TUM,  Technical University of Munich, Munich, Germany

$^3$    University of Padova, Padova, Italy 
\end{center}

\begin{quote} \textbf{Abstract.}{\small 
The Caucasus region is characterized by heterogeneous and strong seismicity as a result of collision between Arabian and Eurasian tectonic plates. A rich variety of seismic events also distinguishes Azerbaijan, located in its south part. In this research, we consider the earthquakes, specifically spatial earthquake patterns that occurred in Azerbaijan and adjacent areas from 2010 to 2023. Applying density-based clustering algorithms to the earthquake catalog, the proper partitions of spatial earthquake distributions were obtained. The statistical properties of the catalog’s partition into 7 clusters are studied in more detail. In particular, we consider the random variable, which is the distance from the fixed point of the earth's surface to earthquake epicenters. The analytical approximation of the cumulative distribution function is constructed for the case when the fixed point coincides with the cluster center and epicenters in the cluster are distributed by the bivariate normal distribution. For comparison,  the numerical distribution functions are evaluated on the basis of Johnson curves and good agreement is observed.
Another case is also considered when the fixed point lies outside of a cluster. Under the assumption that a cluster is a circle and epicenters in it are distributed uniformly, the cumulative and probability distribution functions are derived. Applying these functions to the approximation of histograms for the distances from the Shamkir hydroelectric power station to the clusters shows that satisfactory agreement can be achieved. These results are promising for performing the seismic risk assessment for the Shamkir station or other objects of Azerbaijan’s critical infrastructure.
}
\end{quote}

\begin{quote} \textbf{Keyword:}{\small 
Spatial earthquake distribution, Earthquake clustering, Density-based spatial clustering, Seismicity of Caucasus region.
}
\end{quote}

\vspace{0.5 cm}

\section*{Introduction}
%1,orcid.org/0000-0001-7282-5886
%I.A.Skurativska *1,
%orcid.org/0000-0001-7129-4980
%1,
%orcid.org/0009-0006-8836-2824
%O.M. Sizonenko 2,
%orcid.org/0000-0002-8449-2481
%I.M. Hubar 1,
%orcid.org/0000-0002-2822-7288

\section{Introduction}\label{sec1}

As is known from \cite{stat}, between 50 and 80 earthquakes are registered daily on the Earth. Although a significant portion of them are quite weak and do not pose a significant threat, a few thousand earthquakes each year are still extremely destructive and dangerous. Typically, such strong earthquakes occur near seismic belts, in areas of oceanic ridges, and continental rift zones.

This fully applies to the Caucasus region, where the Greater Caucasus mountain belt is located together with fold-and-thrust belts and a series of depressions \cite{Tsereteli_faults,Martin_Bochud,opentech,Telesca2017,Tibaldi2024}. 
Particularly, the South Caucasus, covering the territories of Azerbaijan, Armenia, and Iran, corresponds to the intercontinental collision zone \cite{Martin_Bochud} and is represented by the Kura, Baiburt-Garabagh-Kaphan and Talysh foreland fold-and-thrust belts \cite{Tibaldi2024}. These zones are seismically active zones connected to the rapid and non-uniform plate convergence between Arabia and Eurasia and exhibit thin-skinned tectonics with fault-bend folds, fault-propagation folds, and duplexes \cite{Tsereteli_faults}. Among the manifestations of active tectonics of the region, we are interested in the set of earthquakes, the properties of which are important for developing models of seismic activity, mechanisms of earthquakes, and forecasting \cite{Tjong-Kie,Telesca2017}, as well as for the effective conduct of economic or other activities, reducing the level of seismic danger, etc.

Since this is a densely populated region, rich in mineral resources (gas, oil, etc.), the study of the Caucasus seismicity and the implementation of earthquake-safe technologies are the subject of many studies \cite{Tsereteli_faults,Martin_Bochud,opentech,Telesca2017}, however, with the development of a seismic station network, by improving the methods of collecting, processing and exchanging seismic data, by developing mathematical and informational methods of data research, such investigations are constantly being deepened and supplemented.

Thus, issues related to identifying groups of earthquakes, i.e. clustering, gained a new impetus for development. The understanding of clustering as a set of aftershocks triggered by the main event was expanded by the interpretation of clustering as the processing of abstract events with the aim of identifying similar features using special cluster analysis algorithms  \cite{Ouillon}.

In particular, this research extends previous investigations of \cite{ICSF24,EGU24} related to the analysis of seismic safety of the territories of Azerbaijan. In studies by \cite{ICSF24,EGU24}, attention was drawn to the research of the Shamkir-Mingachevir reservoir region, where two hydroelectric power stations (HPSs)  are located. Using ground response analysis, local site effects in the vicinity of HPSs were analyzed and recommendations for taking such effects into account were proposed.

This study examines the seismicity of Azerbaijan and surrounding areas and discusses the impact of South Caucasus seismicity on local facilities such as the Shamkir HPS. In particular, using modern approaches in statistical seismology and clustering algorithms, a subset of earthquake epicenters is grouped into {\it spatial}  clusters \cite{Cesca2020,Ouillon,M0eval}, which help to define the shapes of seismogenic zones. The cumulative distribution functions (CDFs) and probability density functions (PDFs) for the distances from a fixed point on the earth's surface to cluster points are constructed. Cases where the distribution functions allow for analytical research are considered, and numerical approximations of the distribution functions are also proposed.
Specifically, by utilizing the relevant distribution functions, the connection of clusters with the Shamkir HPS is analyzed to develop the theoretical foundations of seismic risk assessment.

\section{The earthquakes in South Caucasian  and their clustering}

We consider the South Caucasus region the rectangle enclosed by the Latitude range $38^\circ$ to $42^\circ$ N and the Longitude range $45^\circ$ to $50^\circ$ E incorporating Azerbaijan.

Using the databases of the International Seismological Centre  \cite{earthquakes_bulletin}, the catalog of earthquakes occurred from 01.01.2010 to 31.12.2023. The catalog is quite uneven by year; for instance,  there are about 1200 entries for the years 2010-2013 and about 800 entries for  2014-2017, whereas for  2018-2023, there are already about 4000 entries.
In this research, as in \cite{Telesca2017}, we consider  earthquakes with magnitude equal to or larger than 2 ($M\geq 2$). Depending on the objectives of the study, the lower threshold magnitude can be specified in different ways. One of them is based on the evaluation of the completeness magnitude $M_c$ providing  the seismic catalog completeness. To get a rough estimate for $M_c$,  the MAXC method \cite{M0eval,Telesca2017} can be used. As shown in Appendix \ref{sec:dodatok3}, we can use a rough  estimate  $M_c = 2$ and further consider all earthquakes with $M\geq M_c$.

Next, we are going to deal only with the epicenter coordinates, although certain stages of research can be performed for a more complete list of earthquake characteristics (depth, magnitude, etc.). 
The resulting dataset on the locations of earthquake epicenters from the specified rectangle contains 6097 entries. This dataset is subjected to the clustering procedure, which applies the appropriate clustering algorithms. An overview of this is given in the next section.

\subsection{Survey of  clustering algorithms  and their application to  analysis of seismicity}

Currently, cluster analysis as a tool for the automatic analysis of data of different natures is rapidly developing. Evidence of this is the development of more than a dozen data clustering methods, and the number of clustering algorithms alone already exceeds a hundred.

The most well-known and widely implemented methods include Partitioning Methods (k-means and its modifications  k-medians, k-medoids),  Density-Based Methods  (DBScan and its modifications  HDBScan, OPTICS, ST - DBSCAN - EV \cite{Nicolis}),  Hierarchical Methods 
       (Agglomerative, Divisive), Grid-Based Methods 
        (STING, CLIQUE) and many others \cite{Hennig_book}.
The use of clustering algorithms is designed to reduce the influence of the human factor in data analysis and to automate the same type of work, but the most important thing is to try to identify new regularities and properties in datasets that are difficult to formalize and can be described in fuzzy logic \cite{Ansari2009}.

Therefore, the application of these algorithms to earthquake sequences is more than justified and, as evidenced by the analysis of literary sources, exhibits steady progress. In particular, in the work by \cite{Cesca_cluster}, clustering analysis is applied to moment tensor catalogs, including mining-induced microseismic data. The proposed approach also made it possible to classify earthquake point source models and detect and characterize clusters of focal mechanisms, which potentially allows for the assessment of time-varying hazards in mines. The problems of induced seismicity were also considered in the paper by \cite{Yeck_cluster} when the maximum magnitude earthquake is evaluated caused by the operation of an injection well in Colorado. The developed method incorporates the DBSCAN algorithm for classifying individual clusters and assessing the geometry of individual seismicity clusters. A comparison of the performance of K-means and DBSCAN algorithms with a set of worldwide seismic events that occurred from 1965 to 2016 was carried out  by \cite{Fana_cluster}. The quality of clustering was evaluated based on the comparison of the obtained clusters with known seismic belts. As a result of the research, it was concluded that the DBSCAN algorithm has a significant advantage when working with earthquakes.

\cite{Piegari_cluster} describes the experience of applying the DBSCAN and OPTICS algorithms to analyze earthquake catalogs of 2016 Kumamoto and 2016 Central Italy sequences. The authors carried out extensive analysis concerning the influence of the algorithm parameters on clustering results during explorations of earthquake catalogs. They also concluded that the graphical representation of the spatial distribution of hypocenters and their density helps select the algorithm parameters. Although the selection of input parameters and the type should be made carefully, cluster solutions can provide helpful information about the characteristic structures of a dataset,  features of seismic sequences, and the scenarios of the spatiotemporal evolution of fault systems.

Clustering of earthquakes in the region of the Alto Tiberina Fault system (Central Italy) was also used in the studies of \cite{Taroni}, in which a correlation between clustering and earthquake size distributions was revealed that allows one to assess relations of background and triggered seismicity.

The studies of \cite{Ansari2009} are devoted to identifying seismotectonic provinces in the Iranian Plateau using the fuzzy clustering algorithm. In particular, the spatial distributions of earthquake epicenters were partitioned, and the clustering results were compared with the active faults and seismotectonic models..

Similar problems of detecting the clusters of aftershocks and independent background seismic events were considered in the paper by \cite{Vijay}, where a two-stage clustering approach based on a density-based clustering algorithm was developed. Special attention is paid to analyzing the space-time clustering of an earthquake catalog, including the epicenter plot for California, the Himalayas, Japan, and Sumatra–Andaman.

Using the original ST-DBSCAN-EV algorithm, the authors of \cite{Nicolis} applied cluster analysis to the three large earthquakes in Chile. They identified the precursor seismic activity, the emergence of seismic swarms, and periods of foreshocks and aftershocks. 

Recent studies of \cite{Vallianatos} dealt with the spatial clustering of seismic events forming the Petrinja (Croatia) earthquake sequence. Using the DBSCAN algorithm, several clusters were identified, and their statistical properties in terms of the Tsallis entropy technique were considered.

Due to the rapid development of machine learning tools and clustering methods, in particular, it is obvious that it is not possible to  cover all the current publications in this field of research, even the most significant ones. Therefore, we will conclude the literature review with a few of the most recent works. The (H)DBSCAN algorithms were used in \cite{Mitchinson2024}  for  geospatial and temporal analysis of  seismicity  near Mt. Ruapehu, New Zealand, including the identification of earthquake swarms that occurred in the region. The main feature of this research is the  experience of applying cluster analysis  combined with a comprehensive interpretation of the results  that provides a deeper understanding of the processes in seismically active zones. 
A similar issue related to an earthquake swarm in Italy was considered in  \cite{Essing2024} where the unique approach for automatically processing  clusters of seismicity was developed. The seismicity of  Central Italy 2016-2017 was studied in \cite{Alvarez2024} with the help of the fully  automatic algorithm designed within the framework of the  HDBSCAN, DBSCAN, SOM and  OPTICS cluster algorithms.
Using the clustering technique and neural network, a methodology for  detecting  tectonic tremors among fast earthquakes and anthropogenic events was developed in \cite{Yano2024}.
To revise the seismic catalog of mainland France from 2010 to 2018, in \cite{Grunberg2024}, a new workflow was established  utilizing  the HDBSCAN algorithm.
By applying   the   HDBSCAN algorithm and statistical methods, the basic elements of  spatial clustering and survival analysis of the temporal and spatial patterns for earthquake data were demonstrated   with Python code developed by \cite{Humphrey2024}.

Taking the recommendations into account, particularly those referred to in the references above, regarding the use of algorithms for clustering actual earthquakes \cite{Cesca2020,SCITOVSKI_cluster}, it was decided to apply the DBSCAN and HDBSCAN algorithms in the research.

Let us briefly recall these algorithms' main features \cite{DBSCAN,HDBSCAN,Cesca2020}. Since they belong to the density-based clustering algorithms, the partition of a dataset consists of subsets that differ in density. To evaluate the density near a sample point, we construct the hyper-sphere of radius  $\varepsilon$ (local  $\varepsilon$-neighborhood) and predefine a minimal number of samples ({\it Min\_sample}) to consider the set of points in the hyper-sphere as a cluster. This pair of intrinsic parameters $\varepsilon$ and  {\it Min\_sample} governs the algorithm work \cite{Cesca2020}. All points of the dataset are qualified by core (its $\varepsilon$-neighborhood contains no less than {\it Min\_sample} points including the point of observation), border (it is not the core point but its $\varepsilon$-neighborhood contains at least one core point), and noise points (it is neither a core point nor a border point). The separated cluster is constructed on the basis
of two notions: the density-reachable pair of core points and the density-connected sequence of density-reachable points. The DBSCAN’s main advantages are as follows: does not require pre-definition of a number of clusters; can identify the clusters of arbitrary shapes; uses only two understandable parameters, while its disadvantages \cite{HDBSCAN} include: it is rather sensitive to the variations of $\varepsilon$ \cite{SCITOVSKI_cluster} and  {\it Min\_sample}; may fail when working with the clusters of significantly varying densities; adding new points into the starting dataset requires restarting the algorithm. Other pros and cons can be found elsewhere \cite{hdbscanW}. HDBSCAN \cite{HDBSCAN,hdbscanW} is an advanced version of DBSCAN performing the clustering over varying $\varepsilon$ providing the identification of clusters with varying densities. 

Apart from clustering algorithms, cluster analysis also covers other issues, including the estimation of a number of clusters and intrinsic parameters of algorithms, data preprocessing, dimension reduction for multidimensional data, and cluster validation related to assessment of the clustering quality \cite{Hennig_book}. 

In particular, cluster validation means estimating how well the proposed partition fits the input data and is based on considering the specific measure-validity indices. Various indices stimulate their intensive comparative treatment  \cite{Arbelaitz_index_rev, Todes_index_rev, Liu_validity}.  Here we used the   Silhouette index (or Silhouette score) ($Slh$)   proposed by \cite{Silho_index}.  This index is commonly used in cluster analysis  \cite{DudekSlh,He2024}, including earthquake catalogs \cite{Mitchinson2024,Ibrahim24},  and exhibits acceptable computational complexity and clear interpretation.  Let us provide a short description of $Slh$. 

 When the set $C$ is partitioned by $N $ clusters $C_i$, $i=1,\dots,N$, then for each cluster $C_i$ the Silhouette index  is evaluated as follows
 %5\begin{linenomath}
 \begin{equation}
 Slh(C_i)=\frac{1}{|C_i|}\sum_{j=1}^{|C_i|} s(j),
 \end{equation}
 where $|C_i|$ is the number of elements in the $C_i$ cluster and 
 \begin{equation*}
 s(i)=\frac{a(i)-b(i)}{\max\{a(i),b(i)\}}
 \end{equation*}
 is the sample Silhouette score.  The mean intra-cluster distance $a(i)$ and mean nearest-cluster distance $b(i)$ are calculated by means of the following relations
 \begin{equation*}
  a(i)=\frac{1}{|C_i|-1}\sum_{j=1}^{|C_i|} d(i,j), \qquad b(i)=\min_{k\neq i} \sum_{j=1}^{|C_k|} d(i,j)
 \end{equation*}
 where $d(i,j)$ is a distance (euclidean or cityblock).
 %\end{linenomath}

 By analogy, the  Silhouette score for the set $C$ is the mean value for all $s(i)$, i.e  $Slh(C)=\sum_{i=1}^{N}s(i)/N$.
 
 When all clusters are well separated and dense enough, $Slh(C)$   is close to 1. The low positive $Slh(C)$ value indicates poorly separated clusters. When $-1<Slh(C)<0$, the elements refer to misclassified. 
 Note that the $Slh$ technique works well when clusters are well localized and convex. Moreover, they should be linearly separable. When clusters possess complex shapes and mutual locations do not allow one to separate them by straight lines, then $Slh$ can provide incorrect estimates. Such a situation occurs, for instance, when DBSCAN algorithms are used,  in which clusters are determined by the density of points and can have any geometry. Therefore, a high $Slh$, more than 0.5-0.7, is accepted, which means a confident partition into clusters. A small or negative $Slh$ does not always mean a lousy partition and may require a more detailed analysis of clusters.

 \subsection{Clustering of Earthquake Catalog using DBSCAN algorithm}

To carry out the clustering procedure, we use application packages developed with Python~3.11,  specifically the module {\it sklearn.cluster}   elaborated  within the scikit-learn (ver.1.6) project \cite{scikit-learn}. Python source code implementing the DBSCAN and HDBSCAN (as well as K-means for comparison) algorithms and  calculation  of Silhouette, Calinski-Harabasz ($CH$), Davies-Bouldin ($DB$) indices  is available at  \url{https://github.com/SkurativskaKateryna/AGPH_Earthquake_Clustering_Analysis.git}.

There is no generally accepted methodology to date for estimating parameters for the DBSCAN algorithm \cite{Cesca2020,Piegari_cluster,DBSCAN}. Therefore, we apply a kind of the grid search technique to optimize the partition using the DBSCAN algorithm, which involves testing various discrete combinations of key parameters to identify the best-performing configuration based on the Silhouette score. 
The preliminary estimation for $\varepsilon$ is obtained by applying  the procedure based on the sorted $k$-distance graph. 
This procedure, described in detail in  Appendix \ref{sec:dodatok4}, was implemented taking into account  the recommendations of  \cite{DBSCAN}.

\begin{figure}
\centering
\includegraphics[width=8 cm,height=4.5cm]{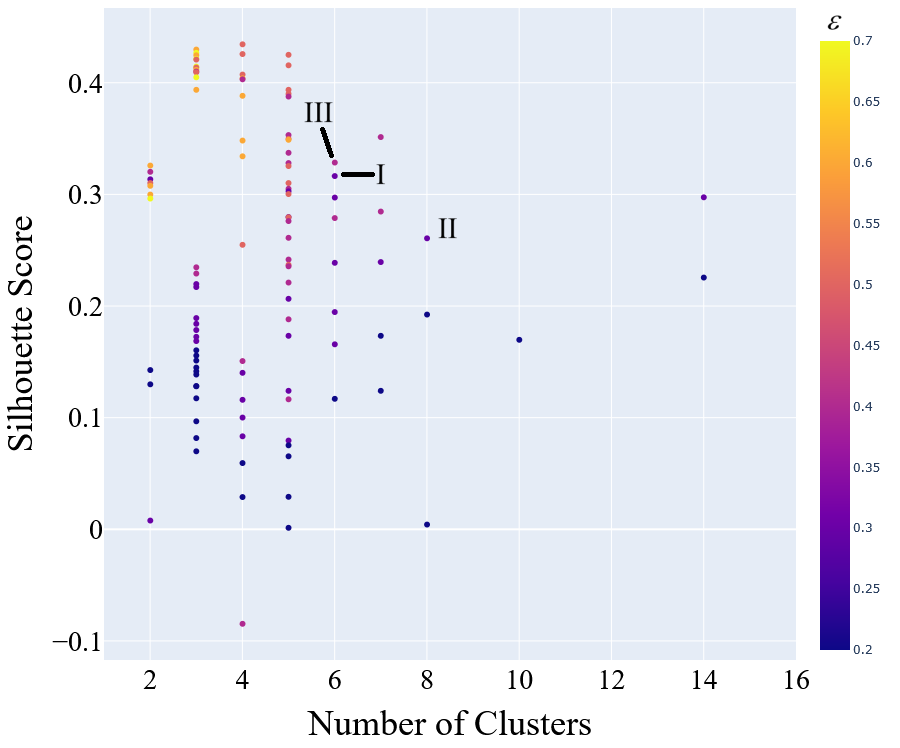}
\hspace{0.1 cm}
\includegraphics[width=8 cm,height=4.5cm]{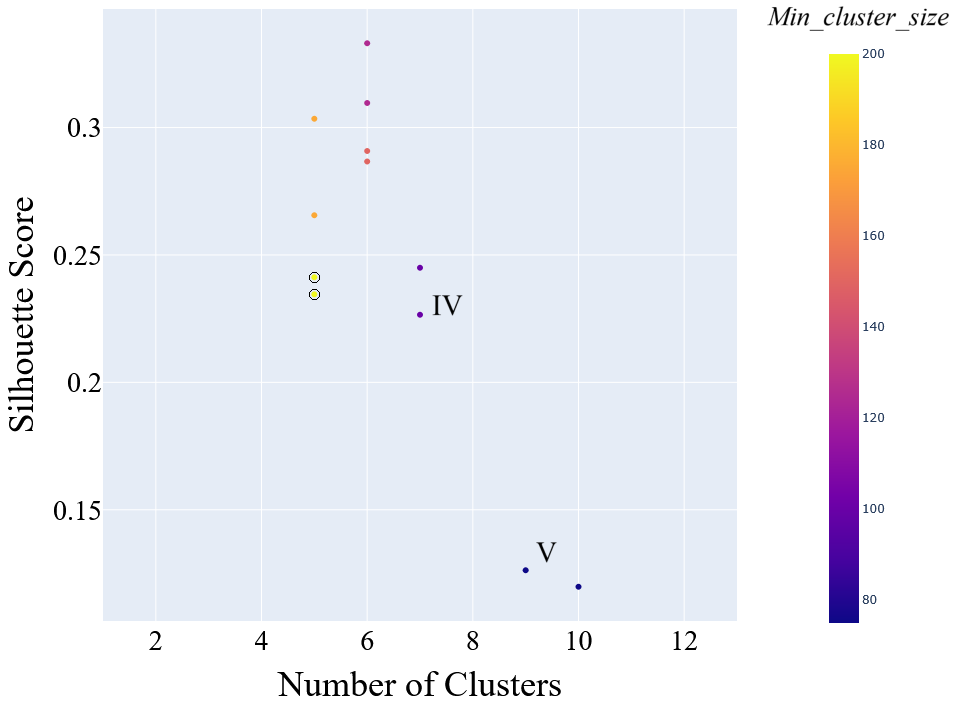}\\
(a)  \hspace{7 cm}  (b) \\
\caption{Silhouette index vs number of clusters evaluated for  DBSCAN   (a) and for HDBSCAN  (b). Roman numerals I -- V stand for the cases considered in the studies.}\label{ds:fig1slh}
\end{figure}

Based on this, we explored a discrete parameter space spanned by the sets:
$\varepsilon \in  [0.2, 1.0]$  with a step of 0.1; $Min\_ samples \in [25; 350]$ with a  step of 25; and 
metric $\in \{euclidean$, $manhattan\}$.
This range was selected to encourage the formation of 5 to 10 meaningful clusters, as cases producing fewer than 2 clusters or more than  about 20 clusters were deemed either uninformative or excessively fragmented. From the total parameter combinations, 113 valid clusterings (with at least two clusters) were retained. For each partition, the Silhouette score was computed to assess clustering quality (Fig.~\ref{ds:fig1slh}a), with similar trends observed when varying either $\varepsilon$ or $Min\_ samples$. Ultimately, we can conclude that large values of both parameters generally produced fewer clusters, no more than 5.

\begin{figure}
\centering
\includegraphics[width=7 cm,height=4.5cm]{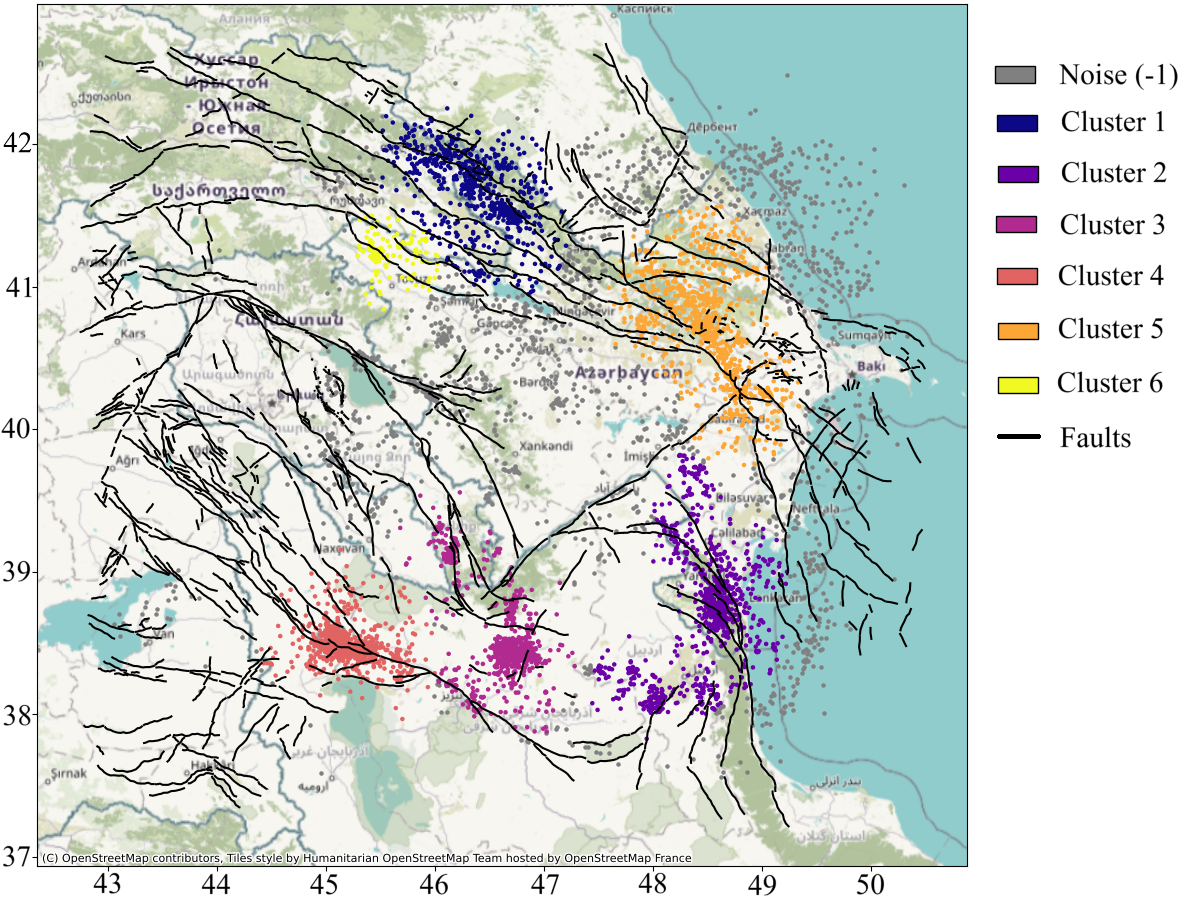}
\hspace{0.5 cm}
\includegraphics[width=7cm,height=4.5cm]{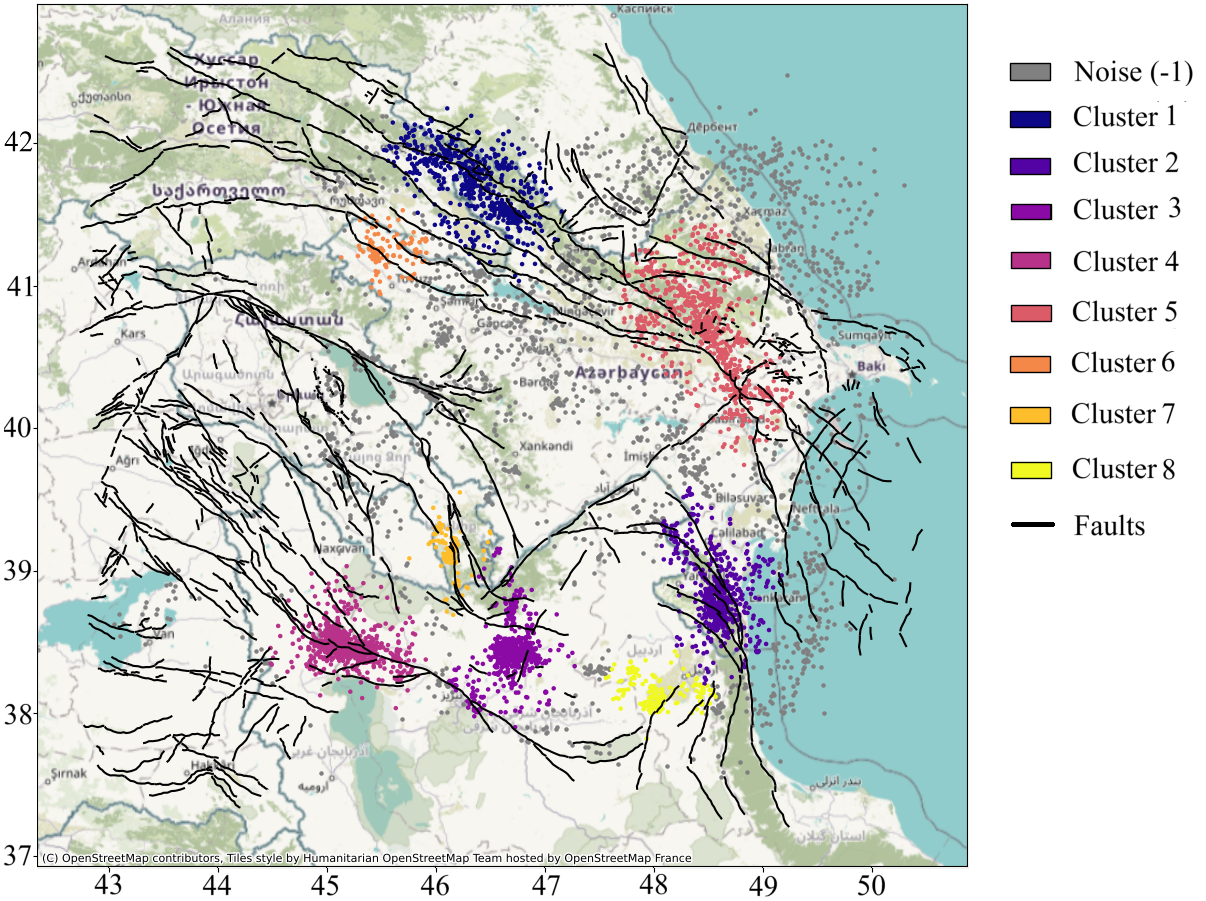}\\
(a) Case I  \hspace{5 cm}  (b) Case II \\
\caption{The  earthquake clustering  by DBSCAN  algorithm. The following groups are marked:   outliers (item Noise $(-1)$ in the legends) and 6 clusters (a) and  8 clusters (b). The fault network is depicted by solid curves.}\label{ds:fig1}
\end{figure}

Instead, let us consider the partitions with 6 to 8 clusters, which may be better suited to territories with an extremely dense fault network. The most features of such partitions can be observed, when  we consider the case I at  $\varepsilon=0.3$, $Min\_ samples =100$,  and metric = ''euclidean''; case II at  $\varepsilon=0.3$, $Min\_ samples =75$, and metric = ``manhattan'' (Fig.~\ref{ds:fig1slh}a). 
The values of Silhouette indices for specified clusters are listed in  Table~\ref{sk:tab1}, where the numbers correspond to the cluster numbers including $(-1)$ standing for outliers  (noise points). All clusters possess acceptable Silhouette indices. The locations of clusters corresponding to the case I are illustrated  in Fig.\ref{ds:fig1}a, 
while those for  case II are shown in Fig.~\ref{ds:fig1}b.

Although we  eliminate   the set of outliers from the further consideration, as was done in many studies \cite{Cesca_cluster,Vallianatos,Yeck_cluster,SCITOVSKI_cluster}, it is still worth noting that they may hold potential  interest for research. Furthermore, these outliers could  correspond to background seismicity, as discussed  by \cite{Mitchinson2024,Essing2024}.

For a more reasonable choice, we offer a division into clusters to compare with the fault network in the Caucasus region \cite{faults_Zel, Avagyan_faults,Tsereteli_faults}. This approach is also used by \cite{Ansari2009,Taroni}.
A comparison of the clustering identified using the Silhouette index with faults is drawn in Fig.~\ref{ds:fig1} and testifies their excellent correlation. In other words, the densest accumulations of earthquakes are identified as clusters, and the location of the clusters correlates quite well with the compaction and intersection of major faults.

{\bf Remark:}
\begin{enumerate}
\item 

 Case III represents the partition into 6 clusters with the higher Silhouette index  ($\varepsilon=0.4$, $Min\_ samples =125$, and metric = ``manhattan''). This partition differs from the case I, however, as the data in Table~\ref{sk:tab1} indicate, this difference is insignificant, at least in terms of Silhouette coefficients.
Formally, to obtain  the partition of the case III, we should consider  the case I, shown in Fig.~\ref{ds:fig1}a,  split Cluster 3  into two clusters, and eliminate Cluster 6, remaining other clusters unchanged, with the possible exception of a few points. Comparing these two partitions, I and III, with the fault network led us to prefer the  partition I (Fig.~\ref{ds:fig1}a).

\item
In addition to the Silhouette index, we also calculated $CH$ and $DB$ indices. Interestingly,  $CH$ indices for the cases I and II reach their maxima among the partitions consisting of 6 and 8 clusters, respectively. This means that corresponding clusters are dense and well separated.  DB indices for the cases  I and II do not reach minimum values, although they do not differ significantly from them.

\end{enumerate}

\begin{table}[h]
\caption{Silhouette index for the clusters obtained by DBSCAN algorithm and shown in Fig.\ref{ds:fig1}}\label{sk:tab1}%
\begin{tabular}{@{}lllllllllll@{}}
\toprule
  & outliers   & 1 & 2& 3& 4& 5& 6& 7 & 8& $Slh(C)$\\
\midrule
 case I & $-0.523 $  & 0.400 & 0.594& 0.642& 0.767 & 0.648 & 0.699 & -- & -- &  0.316 \\
 case II & $-0.568$   & 0.469  & 0.608 &  0.677 & 0.729 & 0.653 & 0.743 & 0.833 & 0.655  & 0.261  \\
 case III & $ -0.507$ & $ 0.748$ & $ 0.622$ & $ 0.648$ & $ 0.702 $ & $0.649$ & $ 0.802$ & -- & -- & 0.323\\
\bottomrule
\end{tabular}
\end{table}

\subsection{Clustering of Earthquake Catalog using HDBSCAN algorithm}

Now we apply the HDBSCAN  algorithm, an advanced version of the DBSCAN, which  constructs a hierarchical density-based cluster tree. This approach enables  more robust and stable identification of clusters, particularly in datasets with varying densities such as earthquake catalogs. Unlike DBSCAN, HDBSCAN does not require the $\varepsilon$ parameter. Instead, it uses other hyperparameters, such as $Min\_cluster\_size$  and a distance metric, as described by \cite{hdbscanW}.
 Other HDBSCAN parameters are used by default according to the  HDBSCAN class description provided by 
\cite{scikit-learn}.

As in the preceding section, we use a similar  technique for choosing the  parameter $ Min\_cluster\_size$,  to identify the partitions consisting of at least six clusters. Thus,  we specify  the range 
 $Min\_cluster\_size \in [75; 200]$ with an increment of  25. The metric used coincides with {\it ``euclidean''} or {\it ``manhattan''}.

The results of the clustering procedure are summarized in Fig.~\ref{ds:fig1slh}b, which shows the relationship between the Silhouette score and the number of clusters obtained for each parameter combination.

Analyzing Fig.~\ref{ds:fig1slh}b, we can see pairs of points.  
These pairs correspond to the partitions with different metrics. It turned out that  they differ insignificantly. As shown in Fig.~\ref{ds:fig1slh}b, an increase in $Min\_cluster\_size$ leads to a decrease in the number of clusters, and a reduction in $Min\_cluster\_size$ leads to an increase.

\begin{table}[h]
\caption{Silhouette index for the clusters obtained by HDBSCAN algorithm and shown in Fig.~\ref{ds:fig2}.}\label{sk:tab2}%
\begin{tabular}{@{}llllllllllll@{}}
\toprule
  & outliers     & 1 & 2& 3& 4& 5& 6& 7& 8& 9 & $Slh(C)$\\
\midrule
case IV & $-0.584$ & 0.741 &  0.724 &  0.819 &   0.745  & 0.667 & 0.808 & 0.734 & -- & -- & 0.226 \\
 case V & $-0.633$   & 0.769  & 0.743 & 0.602 & 0.821 & 0.793 & 0.533 & 0.696 & 0.802 & 0.732 &0.126  \\
\bottomrule
\end{tabular}
\end{table}

\begin{figure}
\centering
\includegraphics[width=7 cm,height=4.5cm]{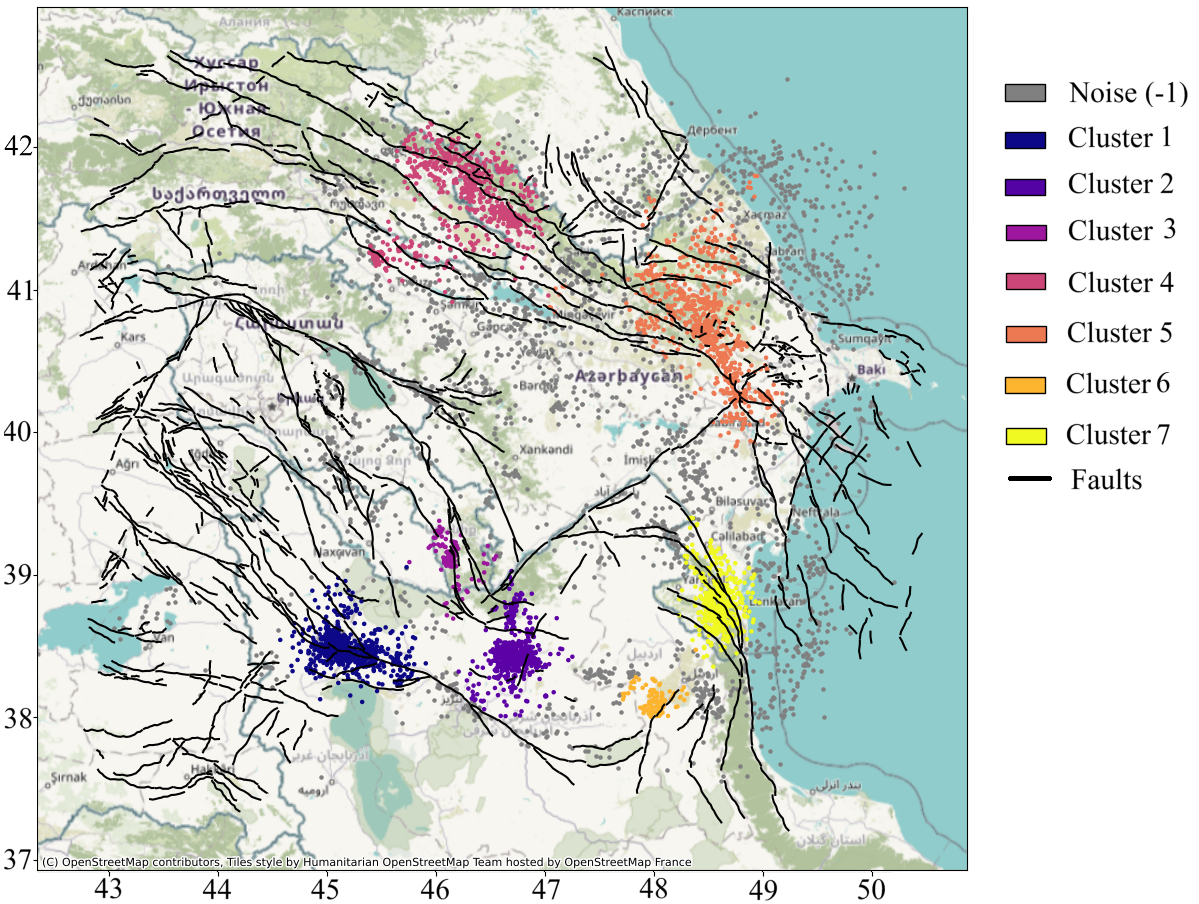}
\hspace{0.5 cm}
\includegraphics[width=7 cm,height=4.5cm]{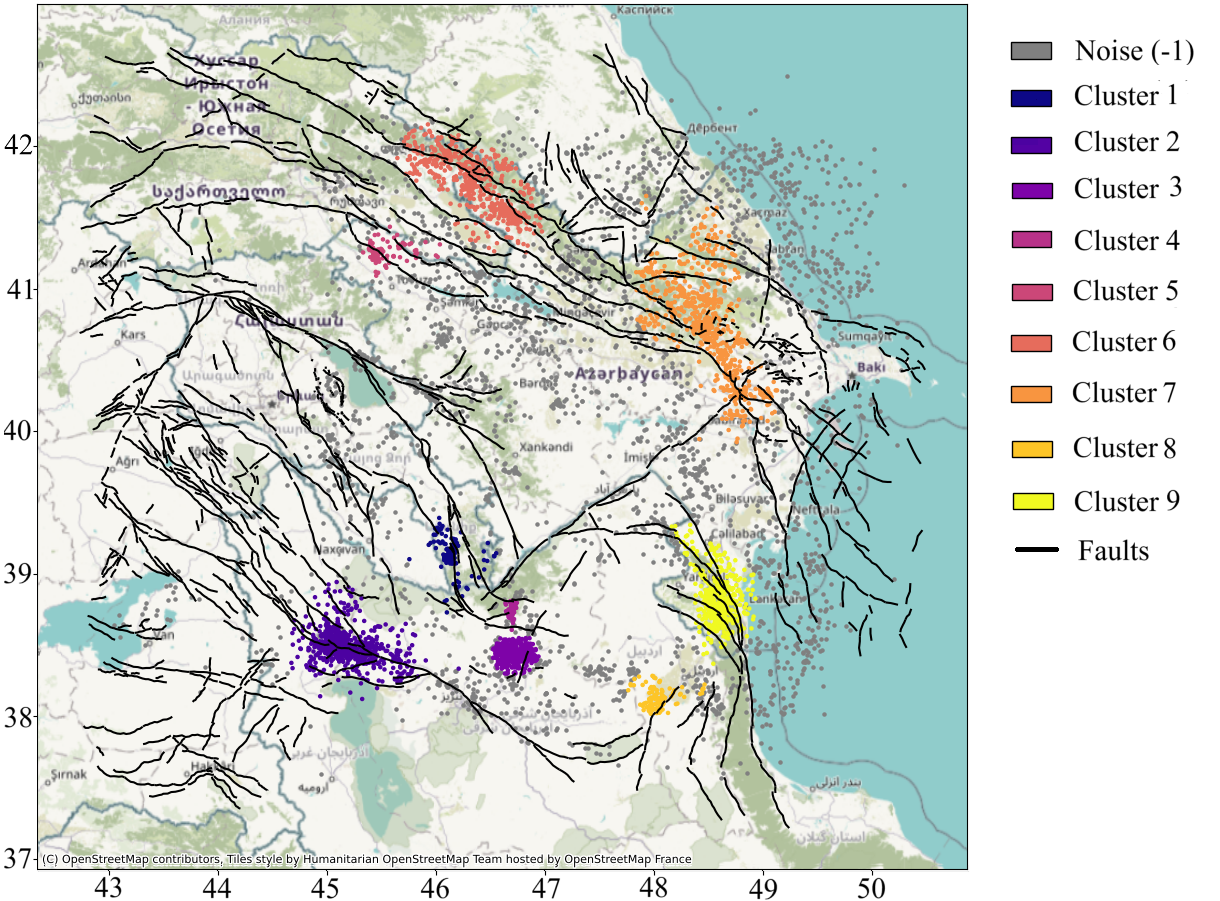}\\
(a) Case IV  \hspace{5 cm}  (b) Case V \\
\caption{The  earthquake clustering  by HDBSCAN  algorithm. The following groups are marked:  7 clusters (a), 9 clusters (b), and the sets of outliers. The fault network is depicted by solid curves. }\label{ds:fig2}
\end{figure}

Comparison with the fault network in combination with taking into account the high Silhouette index shows that the optimal options for partitions can be sets of 7 and 9 clusters. The case IV is observed at $Min\_cluster\_size=100$,  the case V is implemented  at $Min\_cluster\_size=75$. For both cases,  metric = {\it ``manhattan''}.  The Silhouette indices for selected partitions are shown in Table \ref{sk:tab2}, where, as above, the numbers from 1 to 9 mark the cluster numbers, ($-1$) relates to outliers, $Slh(C)$ marks the  Silhouette index for the entire dataset. The data analysis shows that the positive $Slh(C)$  is  0.6 for Fig.~\ref{ds:fig2}a and 0.5 for Fig.~\ref{ds:fig2}b, which indicates a good quality of partitioning. 
 $CH$ and $DB$ indices showed trends rather similar to  the Silhouette score.
 In particular,  the case IV has a $CH$ value  close to the maximum, while  its $DB$ index  reaches a minimum among 7-cluster partitions. 
The location of selected partitions corresponding to the cases IV and V is shown in Fig.~\ref{ds:fig2}a and b, respectively.


\section{Construction of the distance distributions for clusters}

 \begin{figure}
\centering
\includegraphics[width=8 cm,height=6cm]{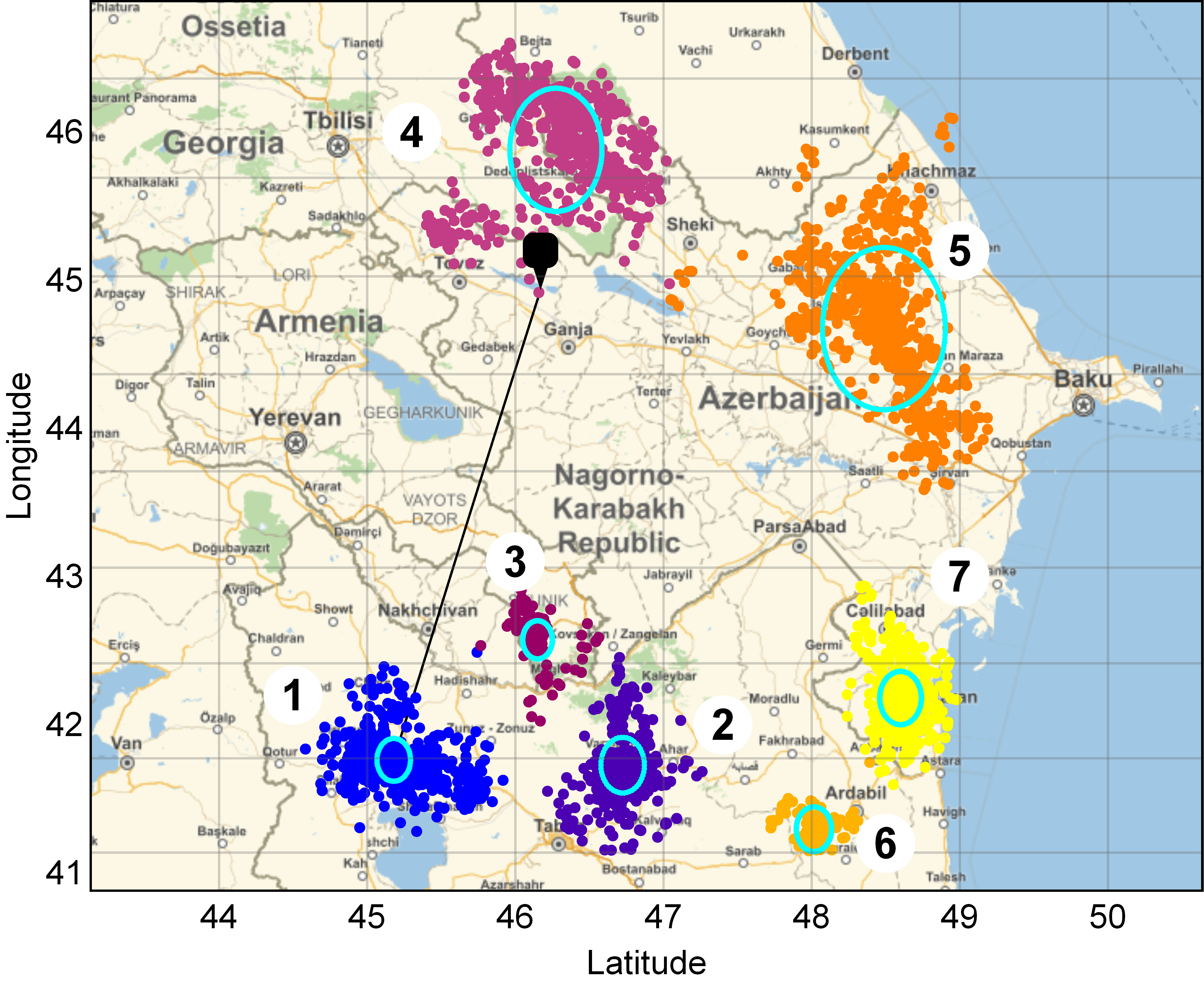}
\caption{The location of the clusters (the numbers 1--7) from Fig.~\ref{ds:fig2}a with dropped outliers. Geomarker corresponds to the Shamkir HPS. The circles bound the domains, for which the CDFs and PDFs are constructed in Sec.~\ref{sk:SecShamkirPDF}.}\label{ds:fig3}
\end{figure}

Constructing the partition among clusters, we identified seismically active zones, which, from the information point of view, are the most significant datasets and contain the most common features. It is further natural to investigate the statistical properties of clusters and the connection of these formations with selected points representing, for instance, strategically essential objects such as the Shamkir HPS. This station is part of a whole conglomerate of objects in this region,  including large Mingachevir, Yenikend, and Shamkir water reservoirs and power stations providing the lion’s share of the power supply to Azerbaijan. This explains this region's extensive geological and geophysical studies   \cite{Tibaldi2024,Babayev_Shem}. 

This research studies the distribution of the random variable, which is the distance from an arbitrary fixed point to cluster points. Note that in the paper of  \cite{Kagan2007} the advantages of such a study method have been emphasized, including independence on system coordinates and grid selection. It should also account for the significant achievements in probability theory and its application in geometry \cite{Kendall,Mathai}.

Thus, let us consider the specified cluster $C$ as a set of random points, defining random vectors from the fixed point $A(a,b)$ to a point  $X(x,y)$ representing
an epicenter of a random earthquake from cluster  $C$. Let the random quantity $W$ be the norm of the random vector
%
%\begin{linenomath}
\begin{equation}\label{sk:dist}
W=||A-X||=\sqrt{\left(x-a\right)^2+\left(y-b\right)^2}.
\end{equation}
%
%\end{linenomath}
 We will further call the vector norm (\ref{sk:dist})  the distance between points, although it is proportional to the traditional distance on the terrain, measured in length units. It is worth noting that it is impossible to carry out comprehensive theoretical studies of this issue in a general formulation due to the limitation of theoretical means. Instead, one can make progress in solving this problem by accepting some auxiliary hypotheses.

\subsection{Distribution of distances from the cluster center to cluster  random points}

Now, we are going to construct the distance distribution when a fixed point coincides with the cluster center located at the point with coordinates   $Q(m_X, m_Y)$, where $m_X$ and $m_Y$ are mean values of the coordinates of the cluster point.   

Let us recall the classical result: if the coordinates of cluster points are independent and normally distributed $N(0,\sigma)$, then the distance (\ref{sk:dist}) has the Rayleigh distribution with the parameter $\sigma$. Therefore, comparing empirical and Rayleigh distributions for the cluster allows one to estimate the correctness of assumptions regarding the distribution of cluster points.

Next, the corresponding CDF for the random variable  $W$ reads as follows 
%\begin{linenomath} 
\begin{equation}\label{skur:distr_fun}
F(z)=P(W<z).
\end{equation}
%\end{linenomath} 
Analytical studies of the function  $F(z)$ can be performed in a few cases. In particular, the analytical expression for $F(z)$  can be derived, as shown in Appendix \ref{sec:dodatok1}, when the coordinates of epicenters of earthquakes are jointly normally distributed and the point of consideration is the cluster center. Such assumptions are quite frequently used in statistical seismology, for instance, during the construction of scattering ellipses for clusters \cite{Ansari2009}, Gaussian-mixture model construction \cite{Aden-An}.

In what follows  the corresponding CDF for the random variable  $W$ reads as follows  for the correlated normally distributed variables $X$ and $Y$, we obtain
%
%\begin{linenomath}
\begin{equation}\label{skur:distr_2Dfun}
F(z)=\frac{1}{\sigma_X\sigma_Y\sqrt{2H}}\int_0^{z^2} \exp\left(-\frac{u(\sigma_x^{-2}+\sigma_y^{-2})}{H}\right)   I_0\left( u \cdot \Delta\right)du,
\end{equation} 
%\end{linenomath}
where $H=2(1-r^2)$ (other designations see in Appendix \ref{sec:dodatok1}).
By definition, PDF  is also described by the following expression
%\begin{linenomath}
\begin{equation}\label{skur:distr_PDF}
PDF(z)=\frac{dF(z)}{dz},
\end{equation} 
%\end{linenomath}
which defines the Rayleigh-like distribution. 

\begin{figure}
\centering
\includegraphics[width=7 cm,height=4.3 cm]{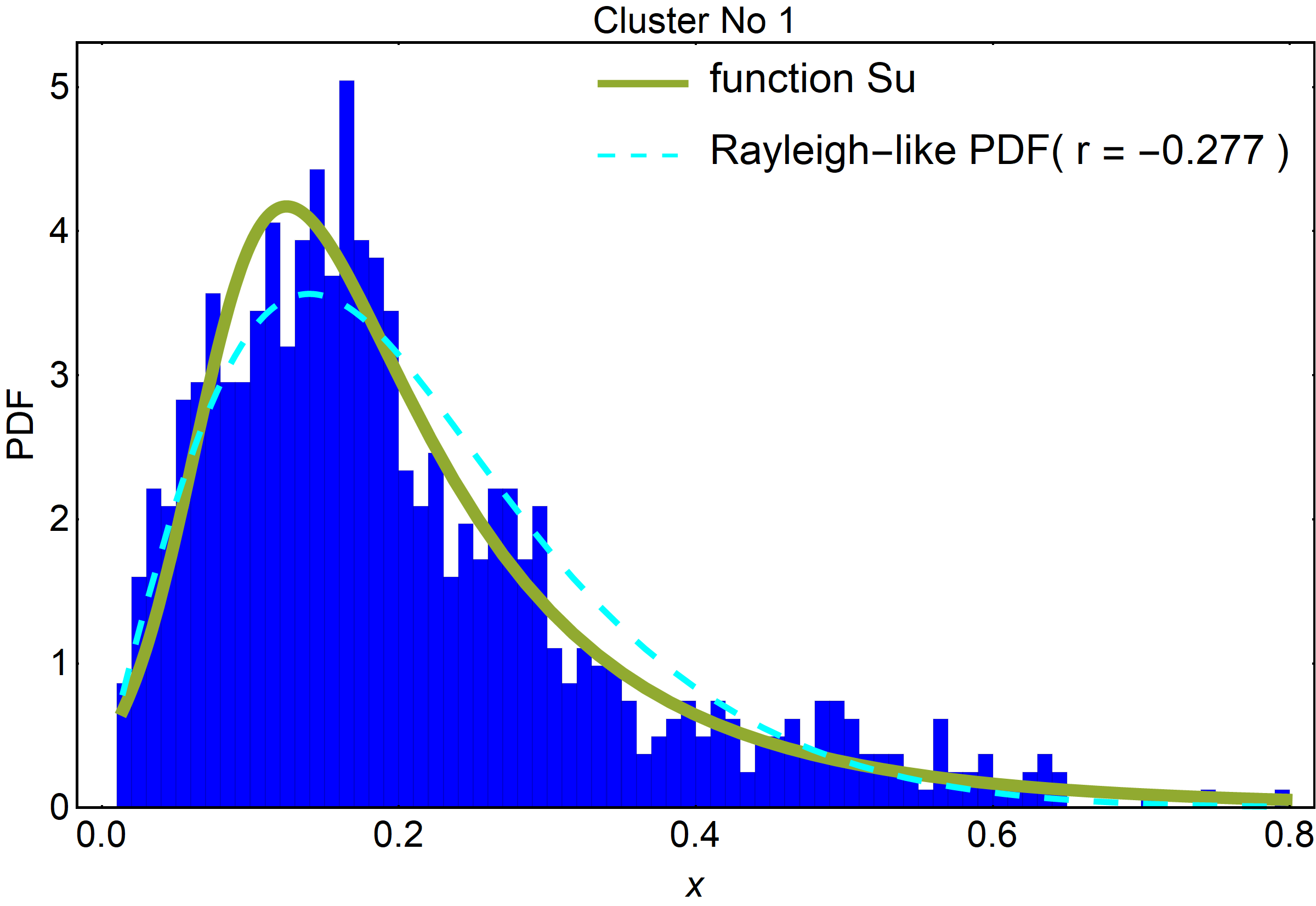}
\hspace{0.5 cm}
\includegraphics[width=7 cm,height=4.3cm]{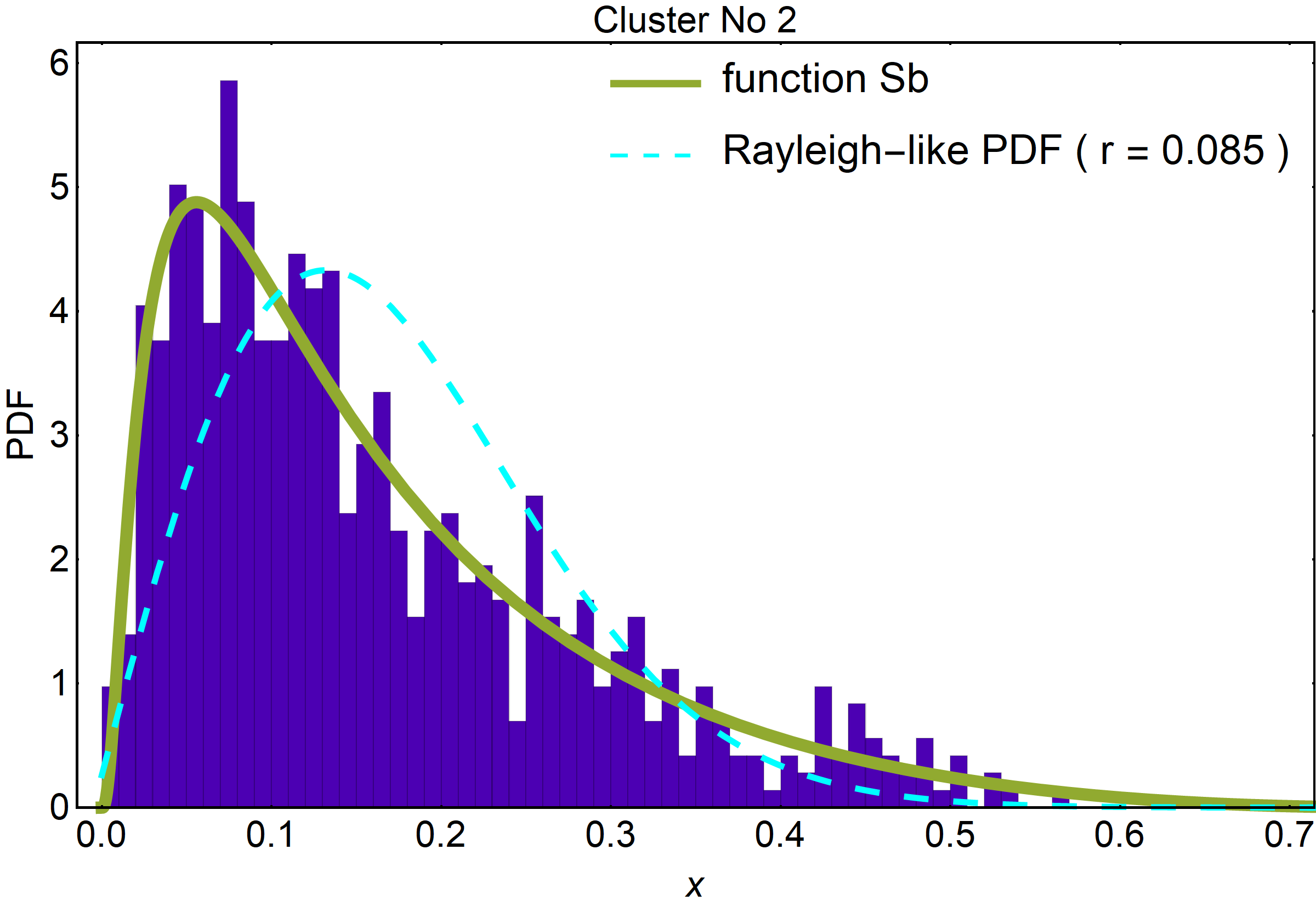}\\
(a)   \hspace{6 cm}  (b)  \\
\includegraphics[width=7 cm,height=4.3cm]{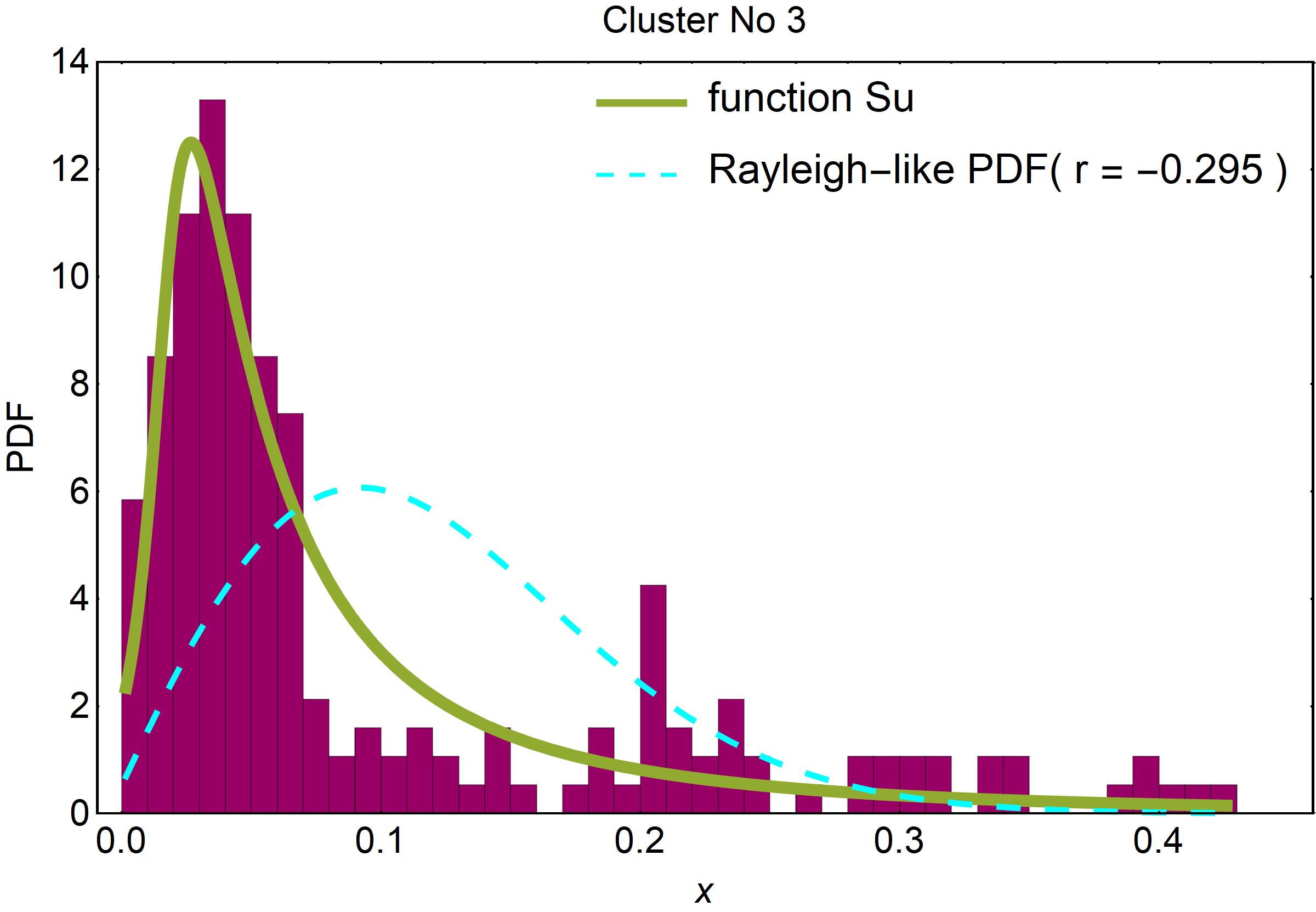}
\hspace{0.5 cm}
\includegraphics[width=7 cm,height=4.3cm]{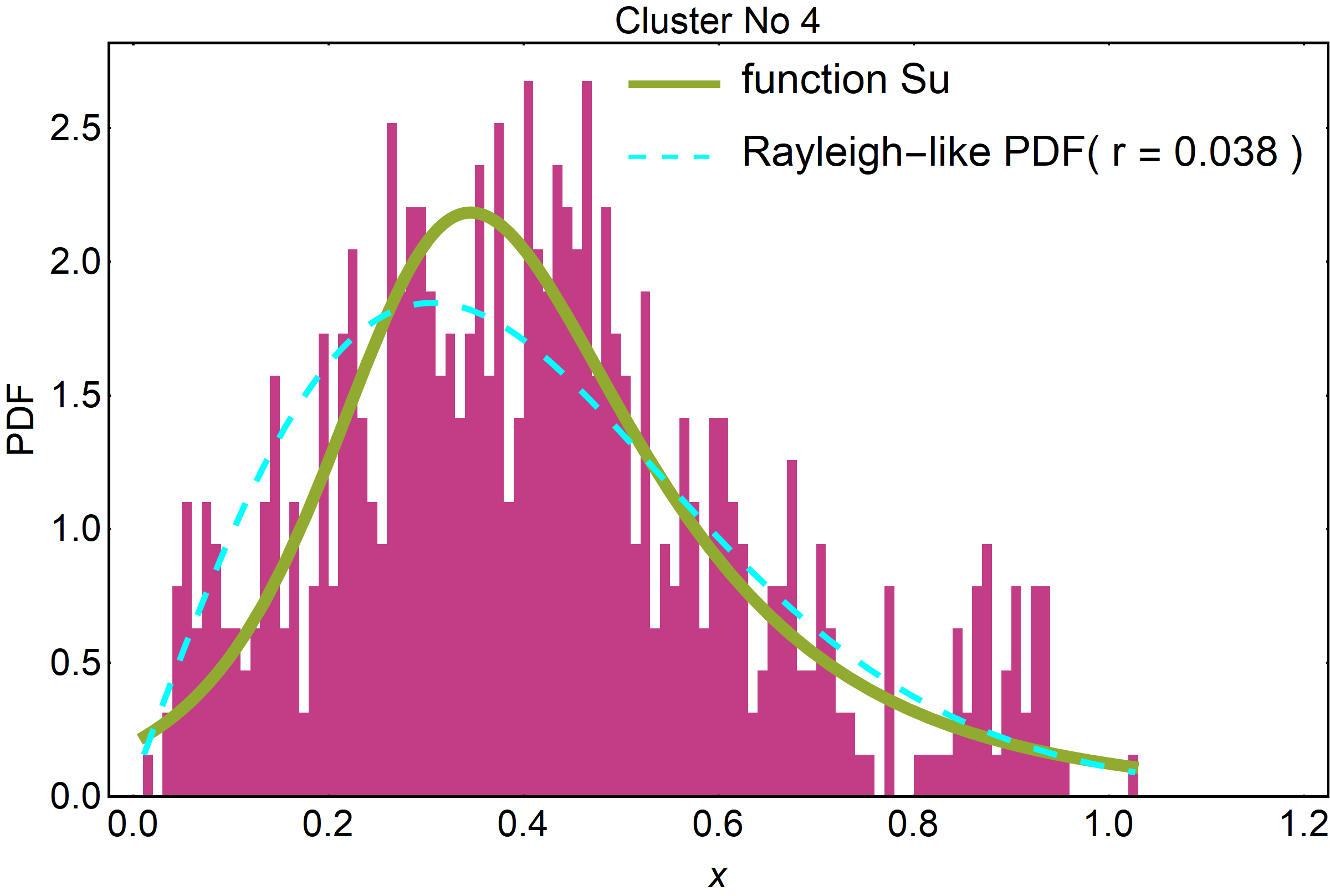}\\
(c)   \hspace{6 cm}  (d)  \\
\includegraphics[width=7 cm,height=4.3cm]{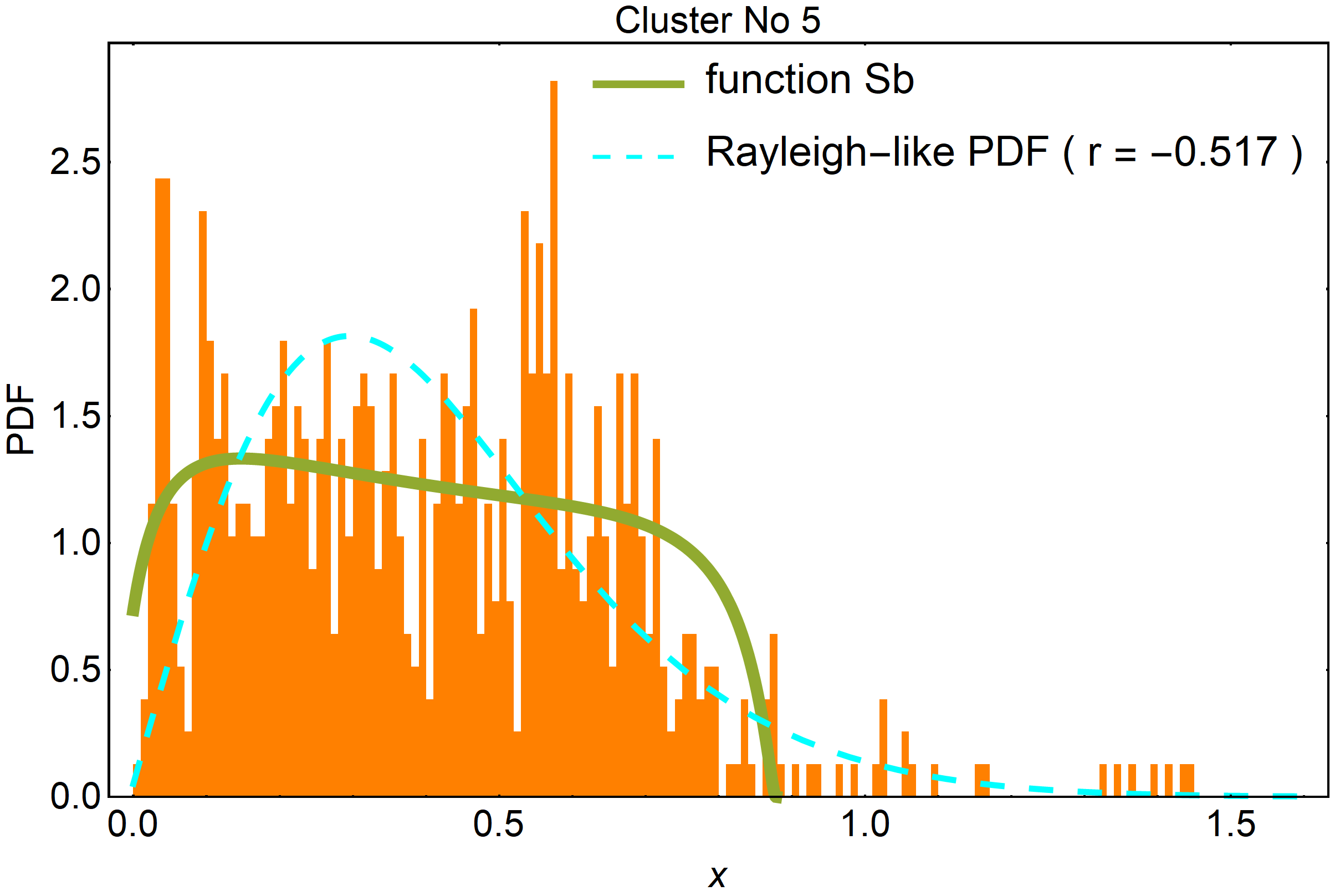}
\hspace{0.5 cm}
\includegraphics[width=7 cm,height=4.3cm]{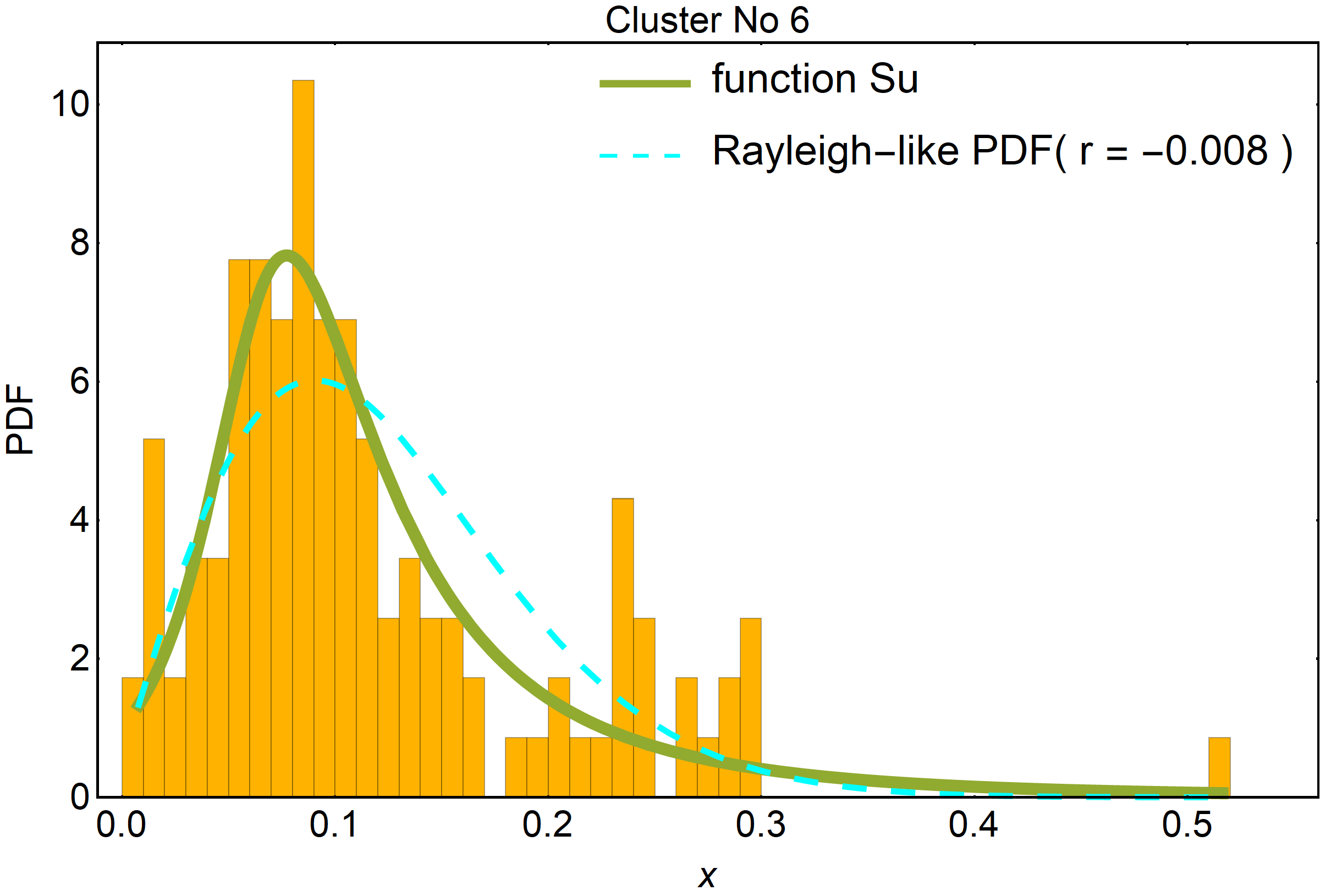}\\
(e)   \hspace{6 cm}  (f)  \\
\includegraphics[width=7 cm,height=4.3cm]{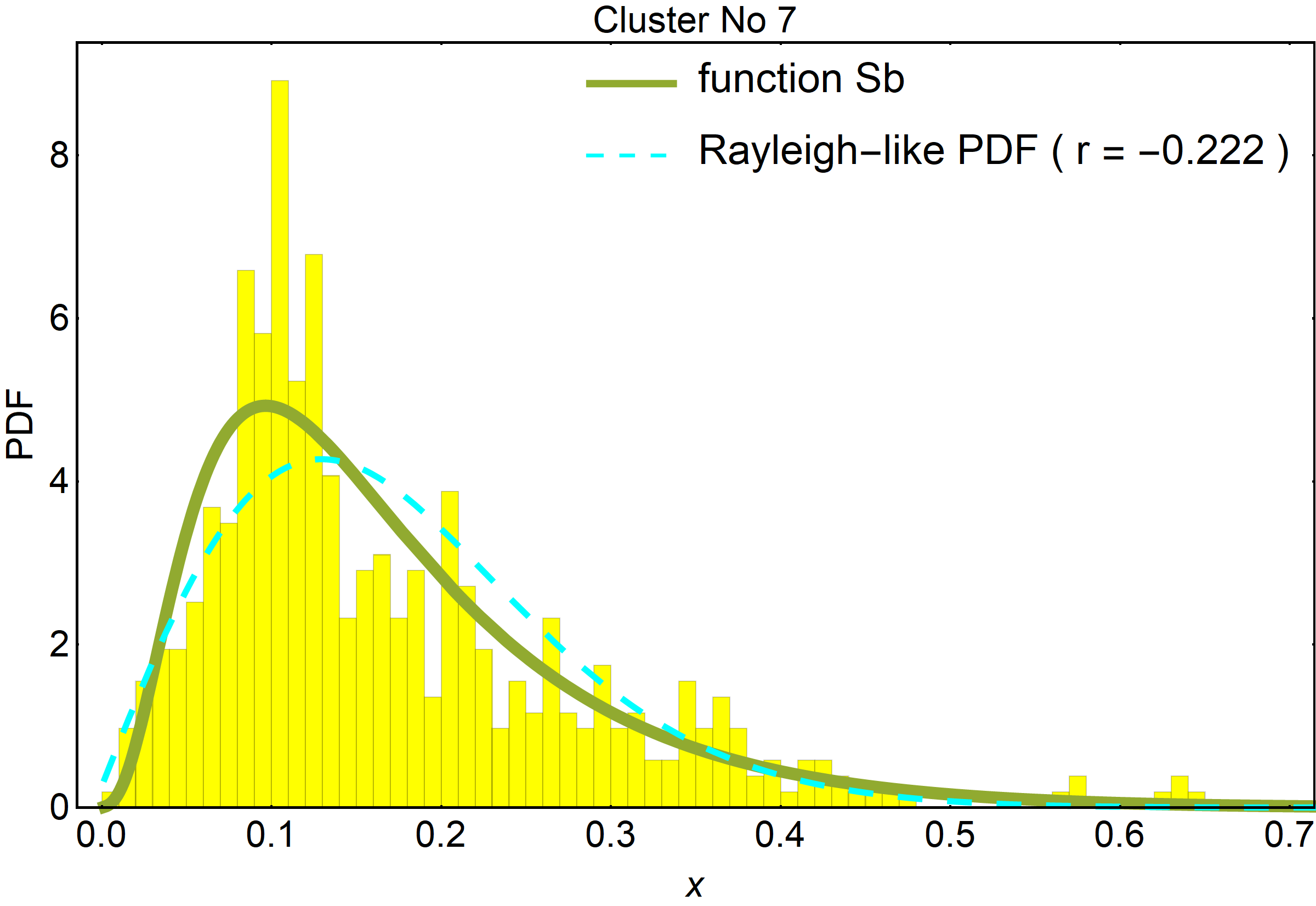}\\
(g)
\caption{The approximation of histograms for the distances from the cluster center to its points. The cluster locations are shown in Fig.\ref{ds:fig3} (the colors of clusters and histograms are the same). Dashed lines mark the  Rayleigh-like distributions defined by  (\ref{skur:distr_PDF}), while solid lines depict the Johnson curves.}\label{ds:fig4}
\end{figure}

Let us consider the functions   (\ref{skur:distr_PDF})  for each cluster of earthquakes under the above assumptions. Figure \ref{ds:fig4} shows the histograms for the variable $W$ defined by  (\ref{sk:dist}). Dashed curves depict the approximations of the histograms by functions (\ref{skur:distr_PDF}).

We also succeeded in approximating the histograms directly using numerical methods. Specifically, we apply the approach based on the PDF approximation by the Johnson curves  \cite{Farnum_JohnsonC, Slifker_JohnsonC}. Step-by-step implementation of the algorithm described in Appendix  \ref{sec:dodatok2} in detail allows one to get pretty good agreement between the approximating curve and the histogram, as seen from the analysis of Fig.\ref{ds:fig4}, where solid lines indicate the Johnson curves.

Note that all approximations fit well with the histograms, except for clusters  3 and 5, whose histograms differ significantly from Rayleigh-like profiles.
One possible reason for the discrepancy between the histogram and the theoretical PDF can be a violation of the conditions for obtaining expression (\ref{skur:distr_PDF}). In particular, the  small number of points in cluster 3 may not be sufficient  to confidently determine the statistical distribution. Another contributing factor could be   a significant deviation of the distribution of points in the cluster from normal. Consequently, as in cases of the histograms for clusters 3 and 5, non-Rayleigh distributions for distances emerge. Such distributions  can be better  approximated by bimodal distributions, representing  a mixture of two unimodal  Rayleigh or normal distributions. A more detailed consideration of this issue requires auxiliary studies.

\subsection{Distribution of distances from the fixed point lying beyond the cluster to cluster random points}\label{sk:SecShamkirPDF}

Now, we consider the case when the fixed point $A(a,b)$ lies outside the cluster. Without restriction of generality,  assume that it coincides with the coordinates of the Shamkir HPS $(a,b)=(40.93, 46.17)$. Its location is depicted in  Fig.\ref{ds:fig3}  by the corresponding geomarker. Let us construct the distribution functions for the random variable $W$, which is a distance from the fixed point $A$ to points of each cluster from Fig.\ref{ds:fig3}.

To conduct theoretical explorations, we should make certain additional assumptions. In particular, we assume that the cluster lies in a closed region. In the simplest case, we consider it to be a circle in which the points are uniformly distributed. This leads us to the classical problem, which admits an analytical solution \cite{Mathai}, p.187.

In particular,  
% 
%\begin{linenomath}
\begin{equation}\label{skur:distr_CDF_Shamkir}
CDF(z)=
\begin{cases}
0,& \text{if $z\leq R-q$,}\\
			\frac{1}{\pi q^2}\left(q^2\left\{\psi-\frac{1}{2}\sin2\psi\right\}+z^2\left\{\phi-\frac{1}{2}\sin2\phi\right\}\right), & \text{if $R-q< z\leq R+q$,}\\
            1. & \text{otherwise,}
		 \end{cases}
\end{equation} 
%\end{linenomath}
and 
%\begin{linenomath}
\begin{equation}\label{skur:distr_PDF_Shamkir}
PDF(z)=
\begin{cases}
			\frac{2z\phi}{\pi q^2}, & \text{if $R-q\leq z\leq R+q$,}\\
            0, & \text{otherwise,}
		 \end{cases}
\end{equation} 
%\end{linenomath}
where 
\begin{equation*}
\begin{split}
\cos \psi&=(-z^2+R^2+q^2)/(2 r R), \qquad 
\cos \phi=(z^2+R^2-q^2)/(2zR), \\
R&=\sqrt{(m_X-a)^2+(m_Y-b)^2}
\end{split}
\end{equation*}
 is the distance from Shamkir HPS to the cluster center and $q$ is the circle radius.

\begin{figure}
\centering
\includegraphics[width=7 cm,height=4.3 cm]{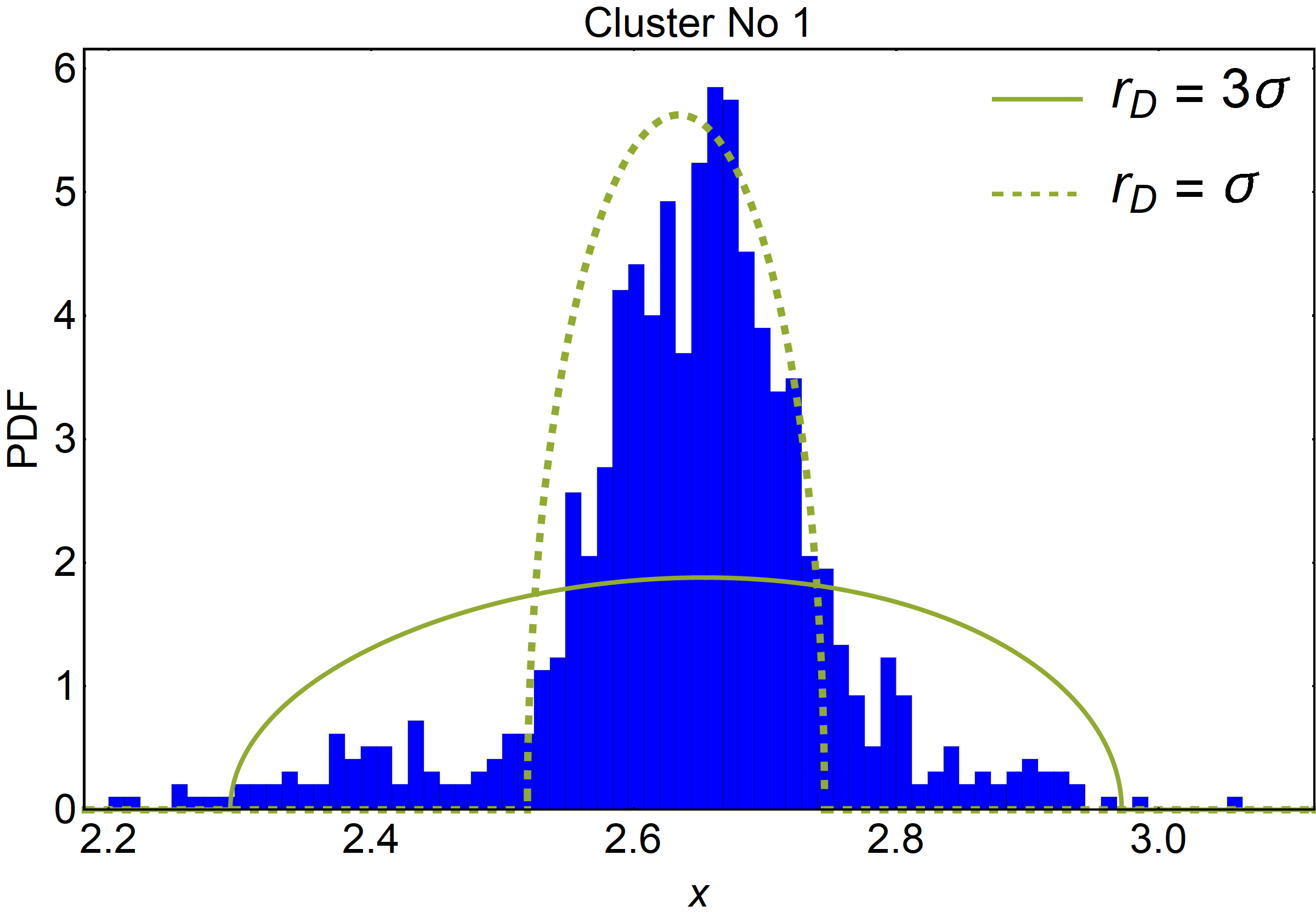}
\hspace{0.5 cm}
\includegraphics[width=7 cm,height=4.3cm]{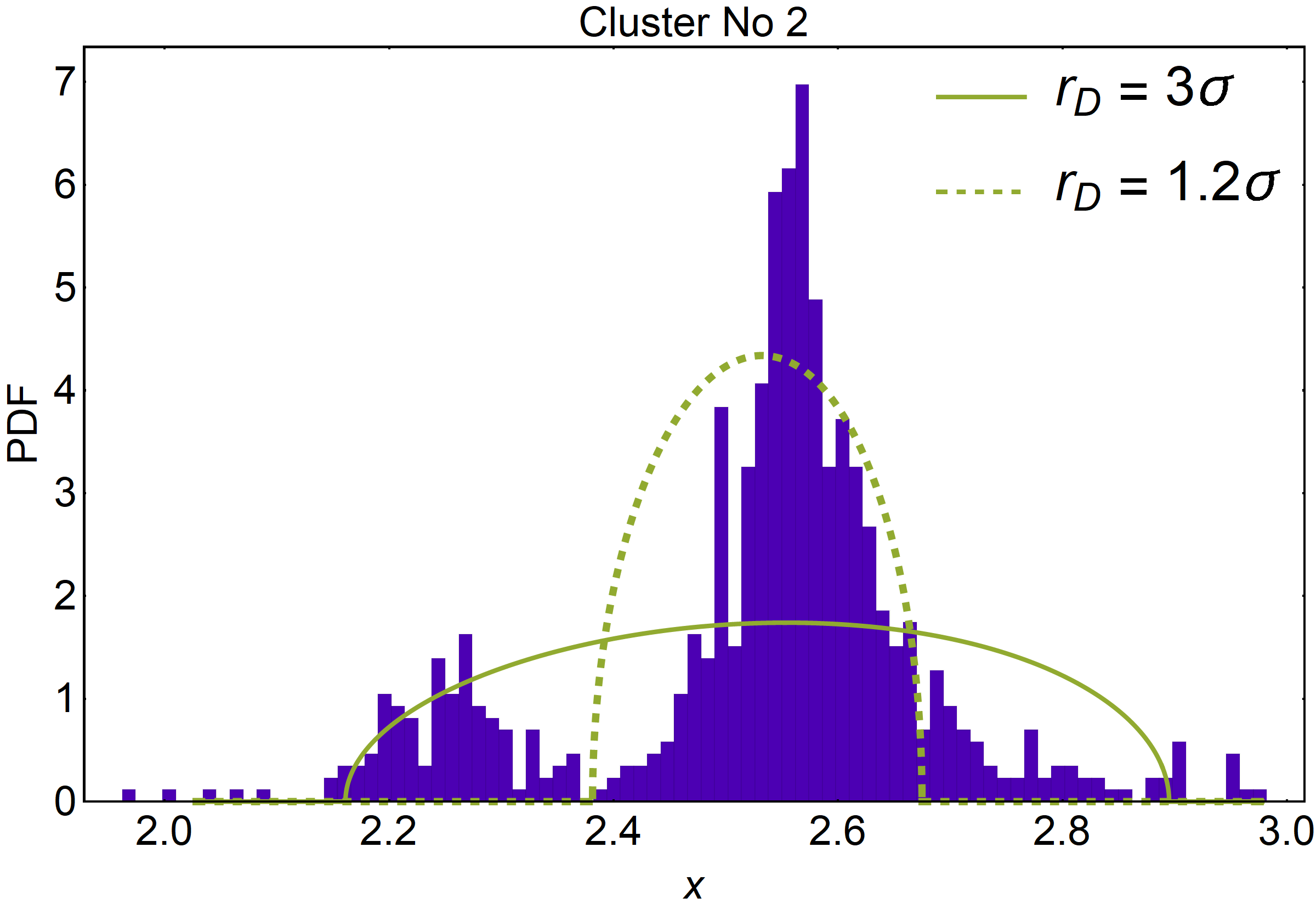}\\
(a)   \hspace{6 cm}  (b)  \\
\includegraphics[width=7 cm,height=4.3cm]{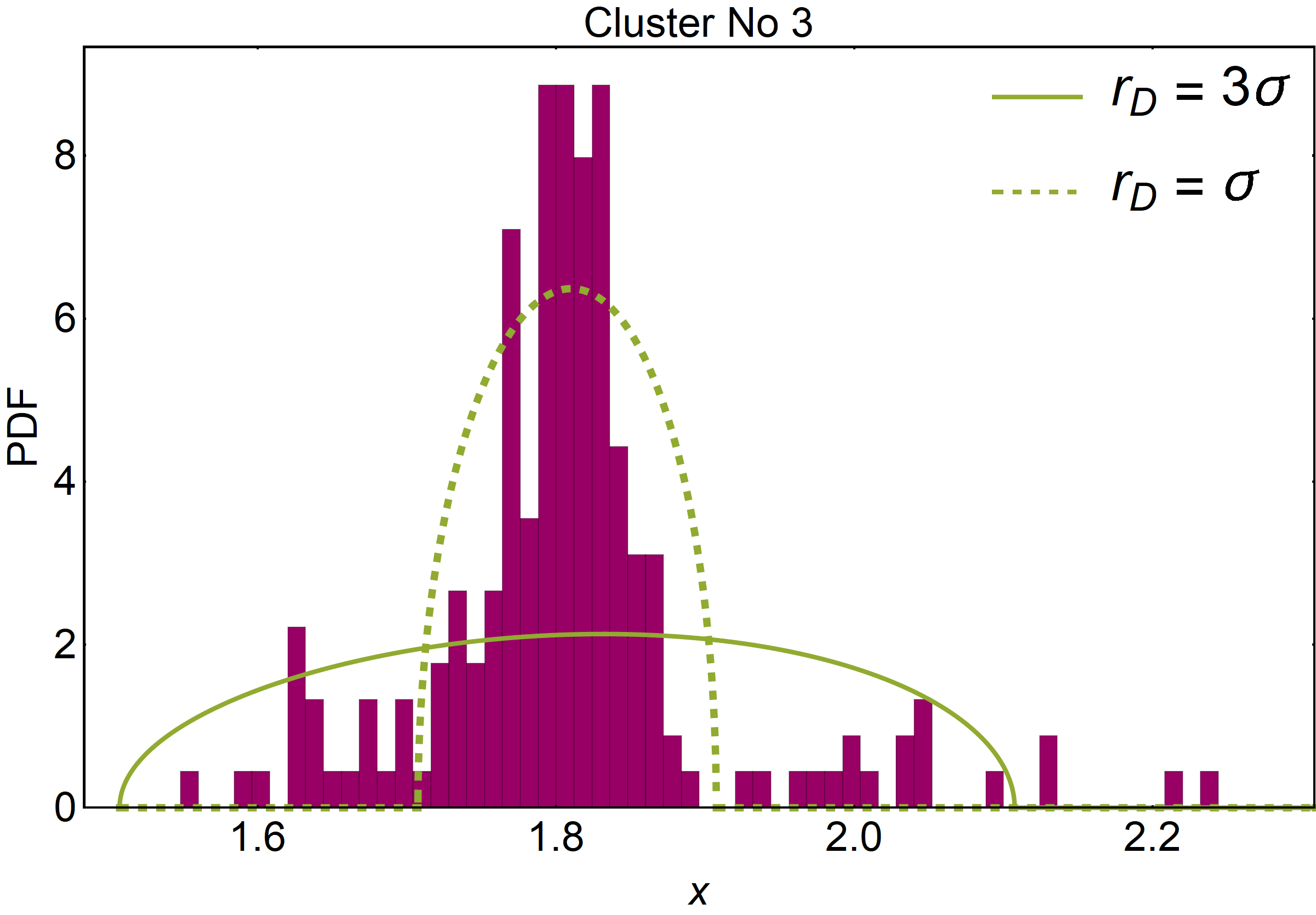}
\hspace{0.5 cm}
\includegraphics[width=7 cm,height=4.3cm]{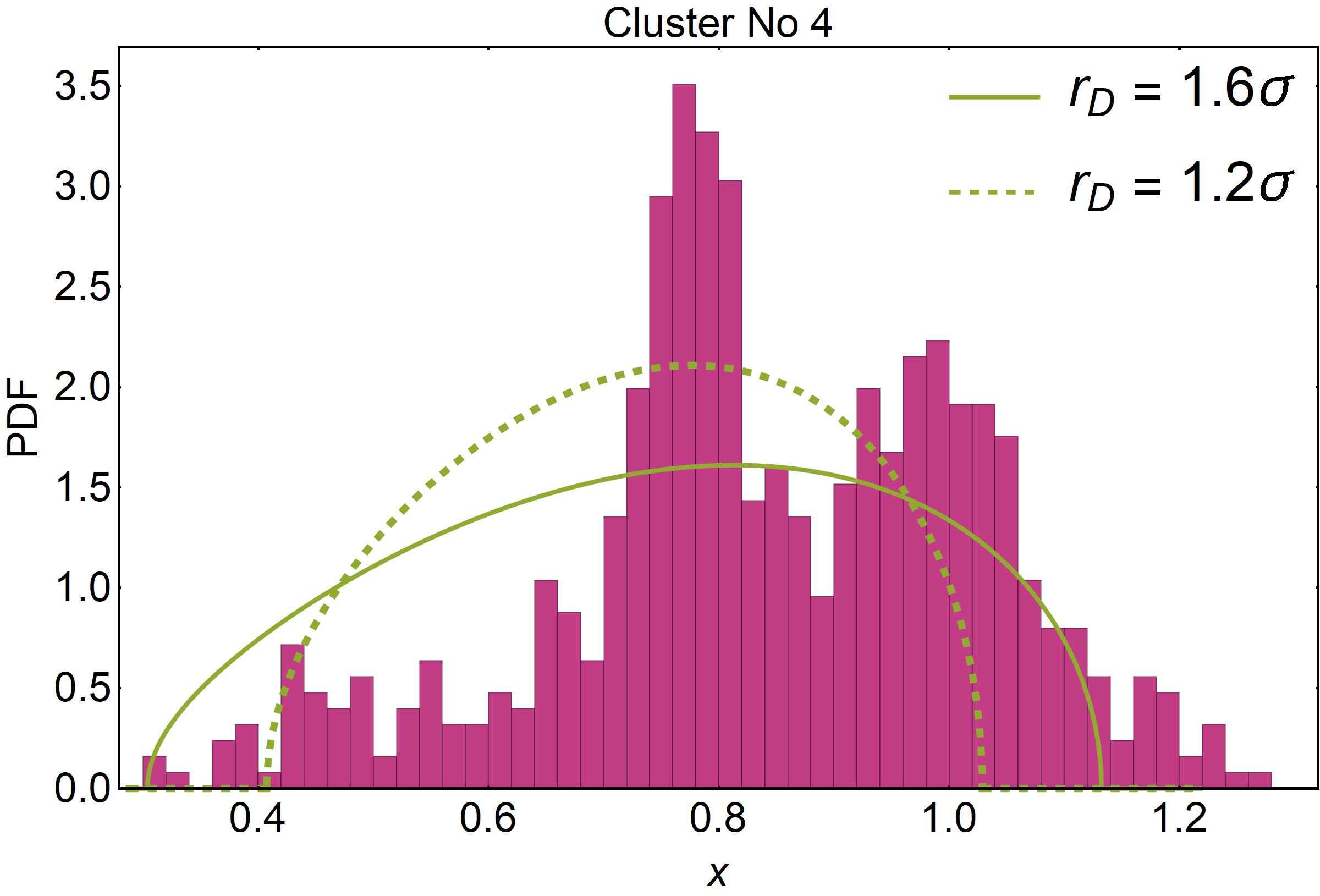}\\
(c)   \hspace{6 cm}  (d)  \\
\includegraphics[width=7 cm,height=4.3cm]{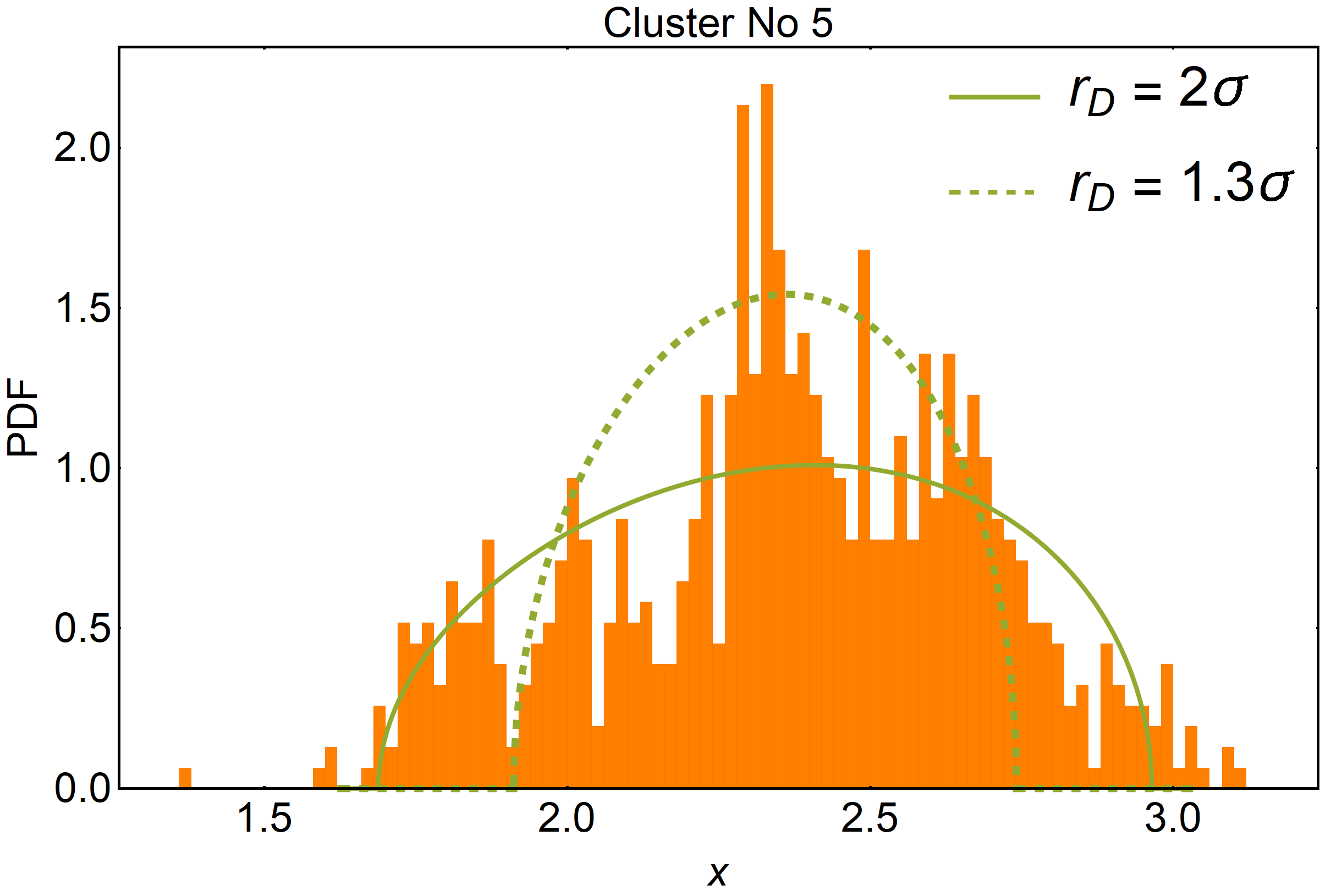}
\hspace{0.5 cm}
\includegraphics[width=7 cm,height=4.3cm]{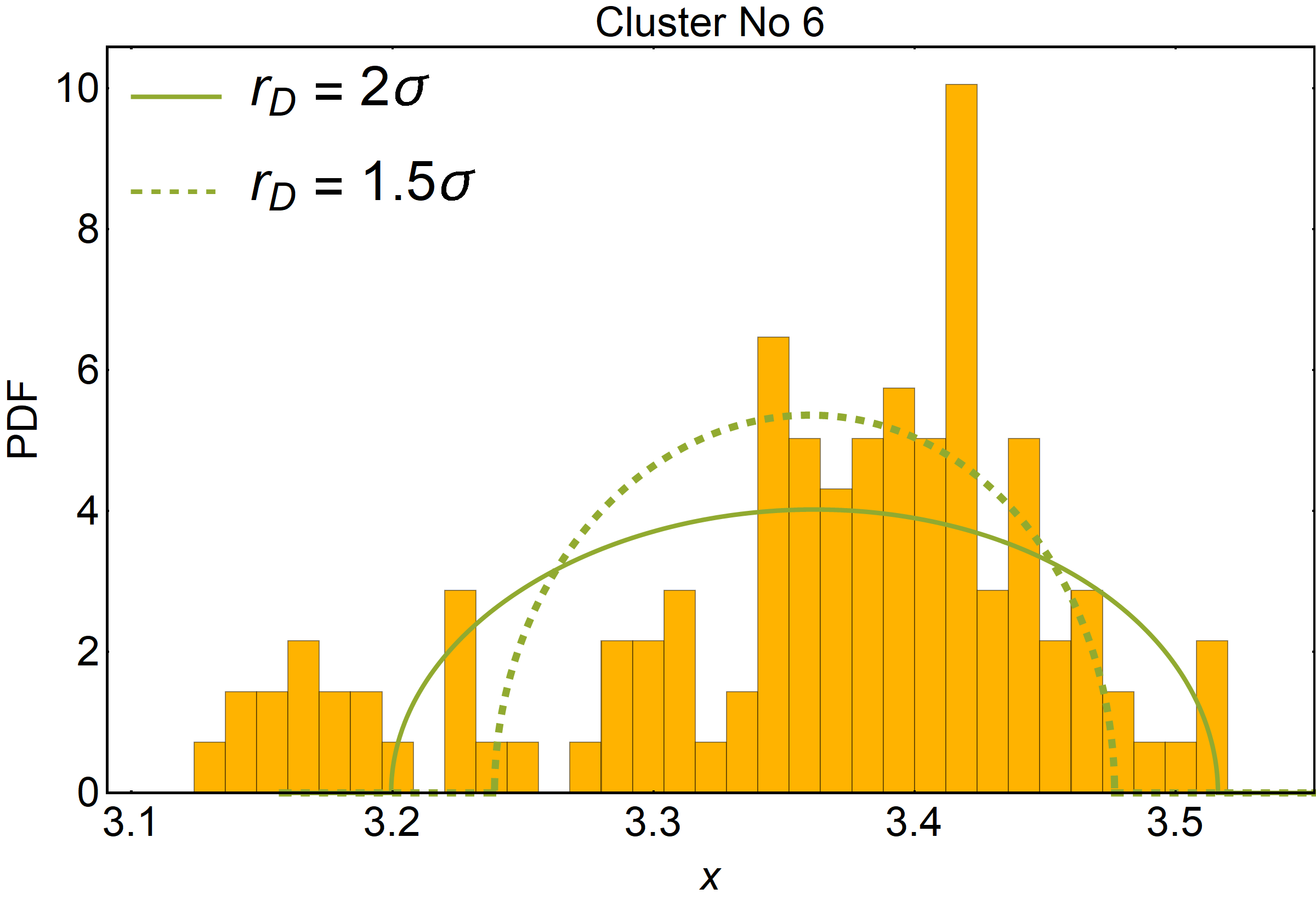}\\
(e)   \hspace{6 cm}  (f)  \\
\includegraphics[width=7 cm,height=4.3cm]{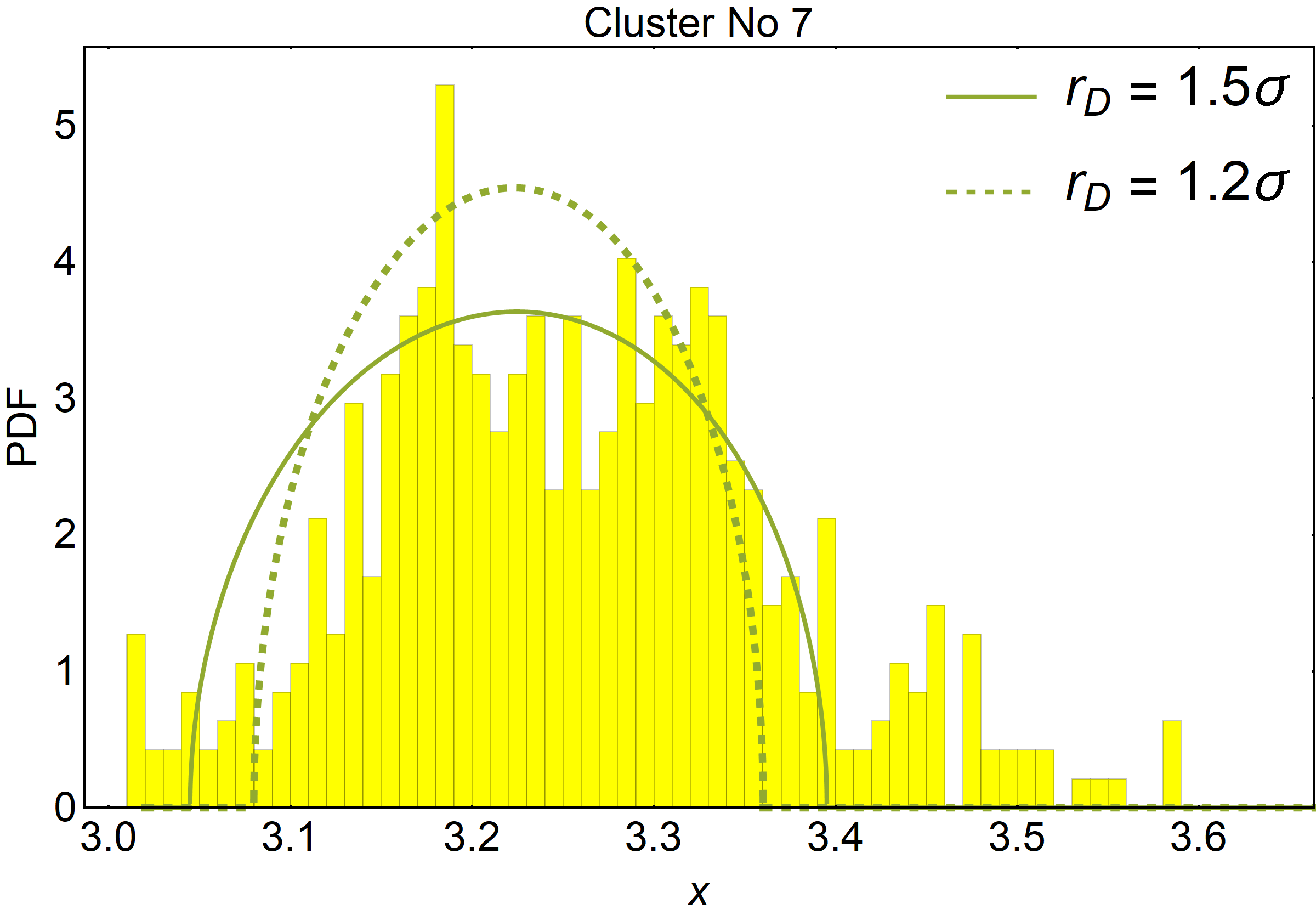}\\
(g)
\caption{The approximations of histograms of distances from the  Shamkir HPS to points of clusters depicted in Fig.\ref{ds:fig3}. }\label{ds:fig5}
\end{figure}

So, let's place the circle center in the cluster's center and choose the circle radius proportional to the standard deviation $\min\{\sigma_X,\sigma_Y\}$. The circle center can be placed at another point, for instance, in (Moda X; Moda Y).
The radius should be small so that the hypothesis of a uniform distribution would be more plausible. Fig.~\ref{ds:fig5} shows a comparison of histograms and corresponding curves at different radii values of circles containing cluster points. An idea of the approximation of clusters by the specified circles can be  obtained from the map in Fig. \ref{ds:fig3}. When constructing the circles, smaller values of $r_D$ were chosen from Fig. \ref{ds:fig5}, i.e., for cluster 1, we take  $r_D=1.0 \min\{\sigma_X,\sigma_Y\}$ etc.

Utilizing the constructed functions (\ref{skur:distr_PDF_Shamkir}) and (\ref{skur:distr_CDF_Shamkir}), one can estimate the probability of the event  $B$ -- the distance from the Shamkir HPS to an earthquake of the specified cluster lies in the interval $R-q<W<R-(2/3)q$. Using the definition of geometric probability, it is easy to understand that the region corresponding to the event $B$ is the lens-shaped domain formed by the intersection of two radii, $R-q$ and $R-(2/3)q$ disks. This domain may be called the near zone of the cluster relative to the fixed external point $A$, which is the Shamkir HPS. Then, we obtain  
%
%\begin{linenomath}
\begin{equation*}%\label{skur:distr_PDF_Shamkir}
P(B)=\int_{R-q}^{R-(2/3)q} PDF(z) dz=CDF(R-(2/3)q).
\end{equation*} 
%\end{linenomath}
%
For the clusters from Fig.~\ref{ds:fig3}, the values of these probabilities are shown in Table~\ref{sk:tab3}.
\begin{table}[h]
\caption{The values of $P(B)$ for 7 clusters}\label{sk:tab3}%
\begin{tabular}{@{}lllllllll@{}}
\toprule
 No. of clrs.  & 1 & 2& 3& 4& 5& 6& 7&  \\
\midrule
$P_i(B)$ & 0.10726 &  0.10697 &  0.10660 &  0.08885  & 0.10211 & 0.10830 &  0.10762  \\
\bottomrule
\end{tabular}
\end{table}
It follows from the analysis of Table~\ref{sk:tab3} that the largest values of $P(B)$ correspond to clusters 1, 2, 3, 6, and 7, located at a greater distance from Shamkir HPS than clusters 4 and 5. This means that the probability of earthquake occurrence in the near zone is higher for the former clusters than that of the latter cluster group. This may be related to the higher density of cluster regions bounded by circles and incorporated in studies.  

Though the obtained findings concern the simplified problem, this problem can help assess the seismic risks of objects, classify territories, and improve insurance compensation algorithms.

\section{Conclusions}

Thus, the studies presented above are concerned with the seismicity of Azerbaijani territories, with a particular focus on the peculiarities of the spatial distribution of earthquake epicenters. Since the region’s seismicity is inherently heterogeneous, analyzing the entire earthquake dataset is not always efficient. This encourages the selection of zones that are more prone to earthquakes and are more seismogenic. Cluster analysis proves helpful in this regard. It is worth noting that cluster analysis operates with numerical data; therefore, the results of its application are not unique and require additional information for decision-making.

Using the DBSCAN and HDBSCAN clustering algorithms, the earthquake catalog was partitioned into clusters. The number of clusters was validated through Silhouette index evaluation  and its alignment with the fault network. 
The derived partitions generally resemble each other, accurately mapping the primary areas of dense epicenter concentration.

Subsequently, the partition consisting of 7 clusters was selected, and its statistical properties were analyzed. Specifically, cumulative and density probability functions were constructed for the random variable representing the distance from a fixed point on the ground surface to the epicenters comprising the clusters in the selected partition. Generally, this problem lacks analytical solutions; however, it can be addressed in two specific cases.

%%%%%

When the fixed point coincides with the cluster center and the epicenters distribution is binormal, the cumulative function can be presented via the integral relation, which effectively approximates the histogram of random distances. In this case, the Johnson curves, which are modifications of Gaussian curves, are evaluated numerically for the histograms. A comparison of theoretical and numerical approximants showed good agreement with the results. Analytical studies are also applicable when the fixed point lies outside the clusters. In this case, we used the solution of the classical problem.

The resulting cumulative distribution function approximates the histograms of distances from the fixed point, representing the location of the Shamkir HPS, to the epicenters. This assumes their uniform distribution over the clusters, whose boundaries were approximated by circles. Although this is a rather rough approximation, the comparison between theoretical PDFs and histograms shows good agreement in the results.

Using the constructed PDFs, the probability of the nearest earthquakes from the selected cluster relative to the Shamkir HPS was evaluated. These findings can help assess the seismic risk for the Shamkir HPS and other strategic objects, including the Mingachevir HPS.

\subsection*{Data availability}  The earthquake catalog used in the current study is available at \url{http://www.isc.ac.uk/iscbulletin/search/}. The fault catalog is available at \url{http://neotec.ginras.ru/index/english/database_eng.html}.
Python source code implementing the cluster analysis of the  earthquake catalog is available at  \url{https://github.com/SkurativskaKateryna/AGPH_Earthquake_Clustering_Analysis.git}.

\begin{appendices}

%\section{Section title of first appendix}\label{secA1}
\section{Calculation of the distribution function for {\it W}}\label{sec:dodatok1}

To construct the CDF for the random variable $W$, we assume that random points in a cluster are distributed by the bivariate normal distribution  
%\begin{linenomath}
\begin{equation*}
f_{XY} =\frac{1}{2\pi\sigma_X\sigma_Y\sqrt{1-r^2}}\exp\left(-\frac{1}{2(1-r^2)}\left[\frac{(x-m_X)^2}{\sigma_X^2}-\frac{2r(x-m_X)(y-m_Y)}{\sigma_X\sigma_Y}+\frac{(y-m_Y)^2}{\sigma_Y^2}\right]\right)
\end{equation*}
%\end{linenomath}
with the correlation coefficient $r$ and the scattering center $(m_X, m_Y)$. Then, by definition, the CDF for $W$ reads as follows: 
\begin{equation}
F(z)=P(W<z)=\iint_G f_{XY} dxdy,
\end{equation}
where 
 $G: \left(x-a\right)^2+\left(y-b\right)^2<z^2$.

Transitioning from cartesian coordinates $(x,y)$ to the polar coordinates $(\rho,\theta)$ by means of the well-known relations
 $x=m_X+\rho \cos\theta$, $y=m_Y+\rho \sin\theta$, one obtains
\begin{equation*}
F(z)=\frac{1}{2\pi \sigma_X\sigma_Y\sqrt{2H}}\int_0^{z^2}  e^{-u(\sigma_X^2+\sigma_Y^2)/(2H)} du \int_0^{2\pi}  \exp\left(\frac{u\left(\sigma_Y^2-\sigma_X^2\right)}{2H} \cos2\theta-\frac{u\cdot  r}{2H\sigma_X\sigma_Y}\sin2\theta \right)d\theta,
\end{equation*}
where $H=2(1-r^2)$.

The integral over  $\theta$ can be written in the following form  
\begin{equation*}
\int_0^{2\pi}  \exp\left[u \Delta \cos(2\theta+\theta_0) \right]d\theta,
\end{equation*}
where $\Delta=\sqrt{\left(\frac{\sigma_Y^2-\sigma_X^2}{2H} \right)^2 +\left(\frac{  r}{2H\sigma_X\sigma_Y}\right)^2}$ and $\theta_0=\arctan \frac{r}{\sigma_X\sigma_Y\left(\sigma_Y^2-\sigma_X^2\right)}$.

 The next change of variable  $2\theta +\theta_0=\eta$ leads this integral to the expression  
 \begin{equation*}
\frac{1}{2}\int_{\theta_0}^{4\pi+\theta_0}  \exp\left[u \Delta \cos\eta \right]d\eta.
\end{equation*}
 
Taking into account the well-known property of integration of  $T$-periodic function, i.e. $\int_a^{T+a}f(t)dt=\int_0^T f(t)dt$, we arrive to  the ultimate result 
   \begin{equation*}
\frac{1}{2}\int_{0}^{4\pi}  \exp\left[u \Delta \cos\eta \right]d\eta=2\pi I_0(u \Delta).
\end{equation*}
 It turned out that the derived integral can be written via the  modified Bessel function of the first kind  
$$I_0(x)=\pi^{-1}\int_0^{\pi} \exp(x\cos\theta)d\theta, \qquad I_0(0)=1.$$ 
It is easy to prove that  $\int_0^{n\pi} \exp(x\cos\theta)d\theta=\pi n I_0$.
This transcendental function as a rule is defined in many packages for technical computing including {\it Mathematica}, where the function  {\sf BesselI[0, x]} is introduced. 

Thus, performing the transformations outlined above, we can present the function  $F(z)$ in the form of the expression 
(\ref{skur:distr_2Dfun}). 

As it follows from  (\ref{skur:distr_2Dfun}),
when the coordinates of a random point are {\it independent} and distributed normally by  $N(0,\sigma)$, i.e. $\sigma_X=\sigma_Y=\sigma$ and $r=0$, then  $W$ is distributed by Rayleigh distribution with PDF 
\begin{equation}
f(z)=\frac{dF}{dz}=\frac{z}{\sigma^2}\exp\left(-\frac{z^2}{2\sigma^2}\right).
\end{equation}
Let us also note that at small  $r$ the function $f(z)$ is close to the  Rayleigh distribution.

\section{Construction of the Johnson curves for the random variable {\it W}}\label{sec:dodatok2}

The algorithm is realized by the tools of the package ``Mathematica''. Let us describe the main stages for the construction of Johnson curves. 

First, we should form the dataset of the quantity $W$ and sort it in ascending order. We also fix, as it has done in the papers  \cite{Farnum_JohnsonC,Slifker_JohnsonC}, the value  
$z=0.524$ and evaluate probabilities 

$P_z$ = {\sf CDF[NormalDistribution[0, 1], z]}=0.699861, 

$P_{3z}$ = {\sf CDF[NormalDistribution[0, 1], 3z]}=0.942025,

 $P_{-z}$ = {\sf CDF[NormalDistribution[0, 1], -z]}=0.300139,

 $P_{-3z}$ = {\sf CDF[NormalDistribution[0, 1], -3z]}=0.0579753, 
 
using the in-buit  cumulative distribution function {\sf CDF[}$\cdot${\sf ]} for the normal distribution $N(0,1)$.

Next, we calculate discriminant $$\Delta=\frac{mn}{p^2}.$$
For 7 clusters from Fig.\ref{ds:fig3}, we have $\Delta =\{$1.28196, 0.713116, 1.4699, 1.62483, 0.394042, 1.9467, 0.89797$\}$.
 Since $\Delta_{1,3,4,6}>1$, we should choose the Johnson $S_u$ curve, while for $\Delta_{2,5,7}<1$ the curve $S_b$ should be chosen.
 These functions (Mathematica syntax) are defined as follows
  \begin{equation}\label{sk:Su}
 S_u(z)= \frac{\eta \lambda}{\sqrt{1+\frac{(z-\epsilon)^2}{\lambda^2}}}
 PDF\left[NormalDistribution[0, 1], 
   \gamma + \eta \mathrm{ArcSinh}\frac{z - \epsilon}{\lambda}\right]
 \end{equation}
 and
 \begin{equation}\label{sk:Sb}
 S_b(z)=\frac{\eta \lambda}{(z - 
      \epsilon)(\lambda + \epsilon - z)} 
 PDF\left[NormalDistribution[0, 1], \gamma + \eta 
     \log\left[\frac{z - \epsilon}{\lambda + \epsilon - z}\right]
     \right].
 \end{equation}

The proper parameter values of each function  can be found in \cite{Farnum_JohnsonC,Slifker_JohnsonC}.

{\bf Remark:} Roughly speaking, the quantity $\Delta_6$ can be considered as close to 1, and then, instead of  $S_b$,  one can use the function $S_l(z )$ defined by the expression 
 \begin{equation*}%\label{sk:Sl}
 S_l(z)=\frac{\eta }{z - 
      \epsilon} 
 PDF\left[NormalDistribution[0, 1], \gamma + \eta 
     \log\left[z - \epsilon\right]
     \right].
 \end{equation*}
It is obvious that $S_l$ is simpler than $S_b$. The profiles of these functions differ insignificantly and approach each other when $\Delta$ tends 1. In this research, we used the functions $S_b$ and $S_u$ only.

\section{Evaluation of the completeness magnitude {\it Mc}  }\label{sec:dodatok3}

\begin{figure}
\centering
\includegraphics[width=7 cm,height=4.5cm]{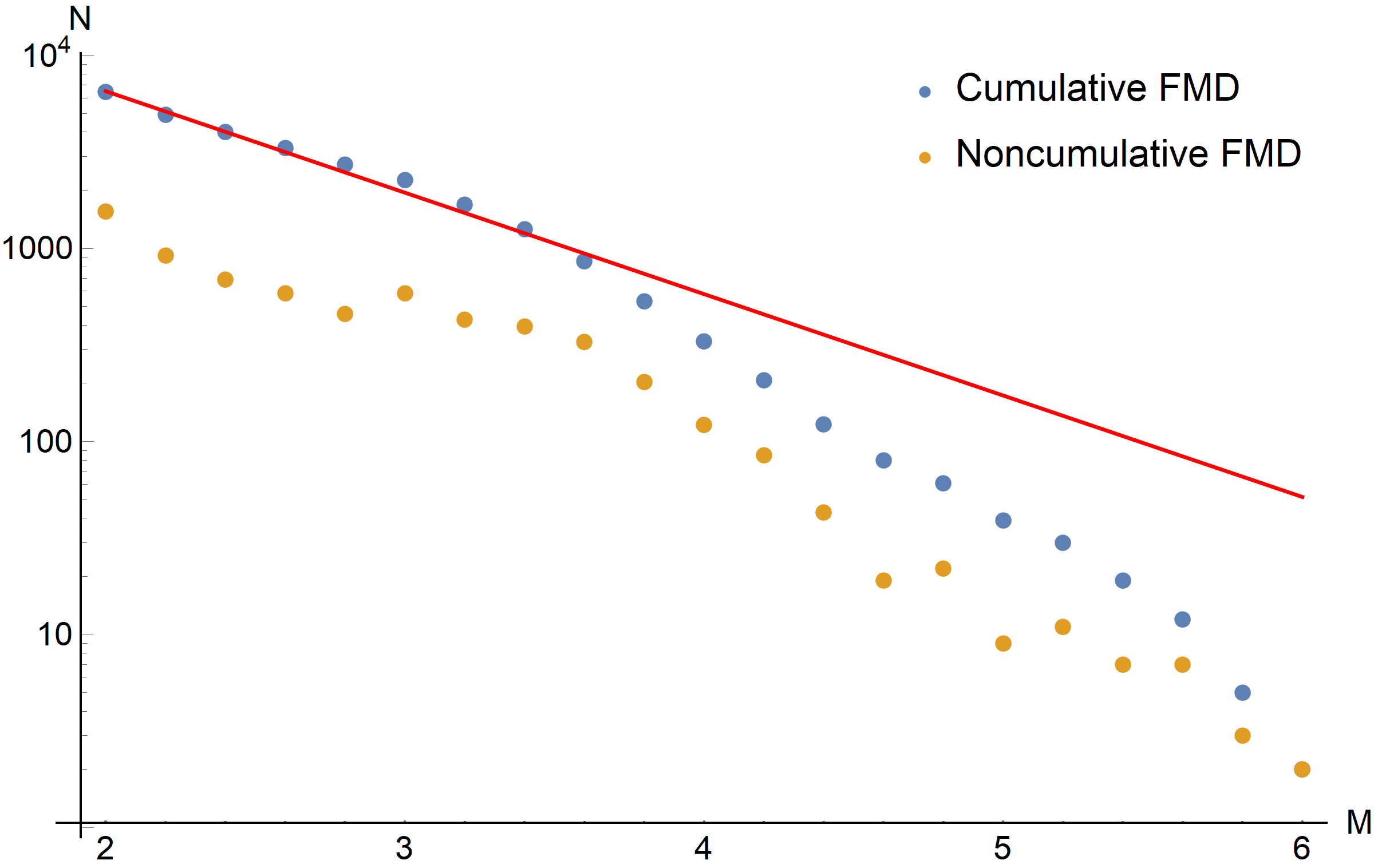}
\caption{Estimation of the completeness magnitude $M_c$ 
using the MAXC  method \cite{Telesca2017}. }\label{ds:M0fig}
\end{figure}

To evaluate the completeness magnitude $M_c$, we consider the cumulative and non-cumulative frequency-magnitude distributions (FMDs) and use the  MAXC  method   \cite{M0eval,Telesca2017,Mitchinson2024}.  The results of calculations are shown in Fig.\ref{ds:M0fig}, where  $M_c=2$ corresponds to  the magnitude for the highest frequency in the noncumulative FMD. The GR law $log_{10} N=a-b M$ with $a= 4.86$ and  $b = 0.52$ (solid straight  line in Fig.\ref{ds:M0fig}) was also  evaluated. 
Note that similar but more comprehensive  evaluations of $M_c$   for  Azerbaijan earthquakes that occurred from 2003 to 2016  
 were performed by  \cite{Telesca2017}, where, in particular,  $M_c=2.1$  was determined  using  the  MAXC  method (see Fig.5a in \cite{Telesca2017}) and corresponding parameters for the GR law $a=4.45$ and $b=0.507$ were calculated. These values are close to our results. Finally, taking into account the fact that $M_c$ usually decreases  with time in most catalogs \cite{M0eval} and exhibits significant spatial  heterogeneity, varying   by an order of magnitude \cite{M0eval}, we  can conclude that the completeness magnitude $M_c=2$ for our studies is quite appropriate estimate for the lower magnitude threshold  of the earthquake catalog used in our study.

\section{ The parameter  estimation for DBSCAN}\label{sec:dodatok4}

\begin{figure}[h]
\centering
\includegraphics[width=7 cm,height=4.5cm]{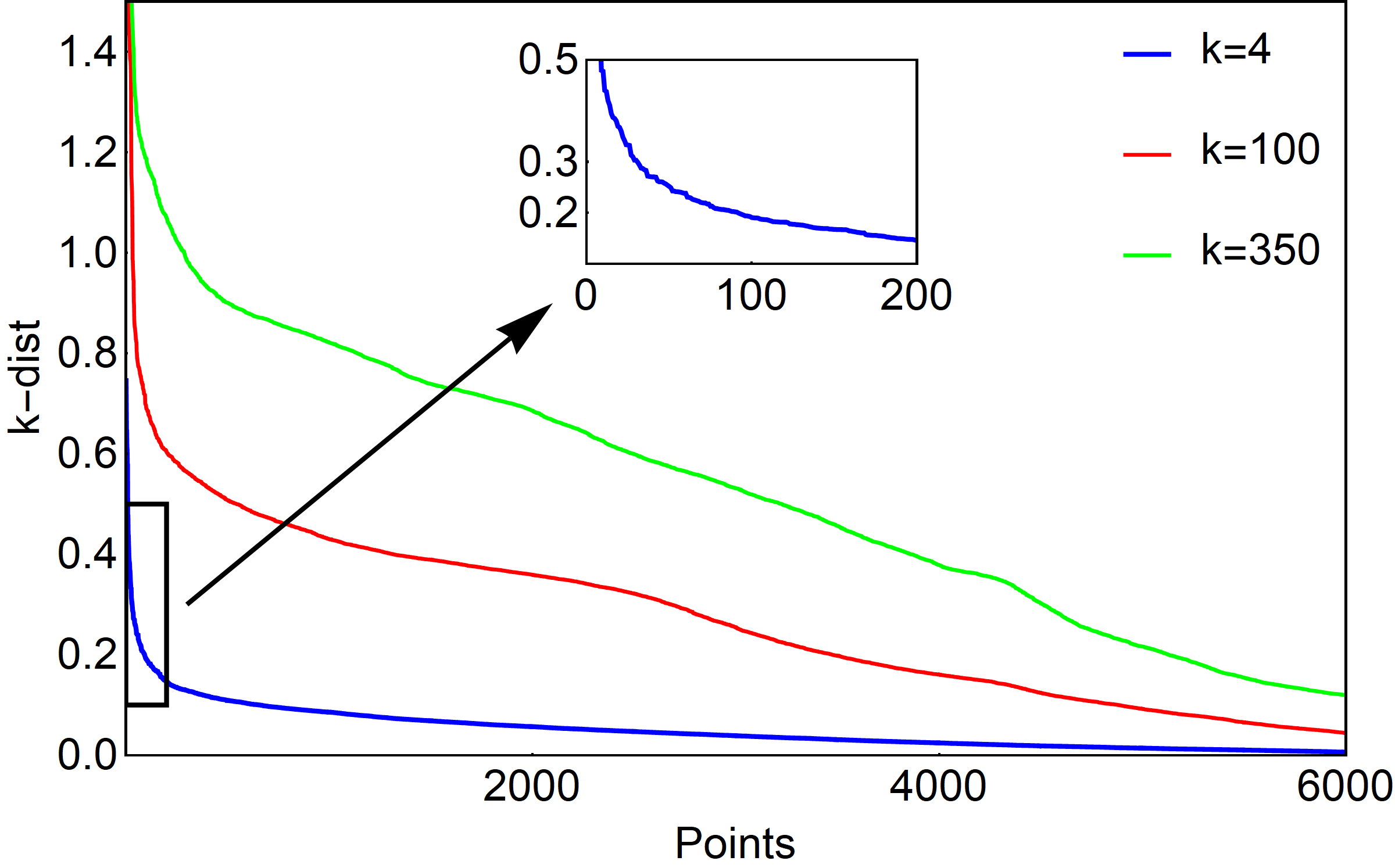}
\caption{The sorted $k$-distance graphs. }\label{ds:kdist}
\end{figure}

To select the parameter $\varepsilon $  and  $Min\_samples$ for the DBSCAN algorithm, let us consider the sorted $k$-distance graph mentioned in the papers \cite{DBSCAN,Fana_cluster}, using the sorted distance matrix. We start from plotting the $4$-distance graph (Fig.~\ref{ds:kdist}), where the $\varepsilon $ value relates to the ``knee''  formation on the graph.  
 Analysis of $k$-distance graphs obtained at $k=100$  and $k=350$ shows that the  corresponding ``knee'' value of $\varepsilon$ increases, as $k$ increases. However, as noted in  \cite{Piegari_cluster,Cesca2020}, this procedure should still be considered as an approximate method for parameter estimation.

\end{appendices}


\bibliographystyle{IEEEtran_link}

\bibliography{skur_2Dcluster}% common bib file

\end{document}